%% file: article.tex
\documentclass[preprint]{aastex}
\usepackage{times,graphics,rotate,psfig,epsfig}
\newcommand{\SiIII}{Si~{\sc iii}}
\newcommand{\SiII}{Si~{\sc ii}}
\newcommand{\SiIV}{Si~{\sc iv}}
\newcommand{\HII}{H~{\sc ii}}
\newcommand{\teff}{$T_{\rm eff}$}
\newcommand{\Ha}{H$_\alpha$}
\newcommand{\Hg}{H$_\gamma$}
\newcommand{\eps}{$\epsilon$}
\newcommand{\g}{$\log g$}
\newcommand{\CII}{C~{\sc ii}}
\newcommand{\OII}{O~{\sc ii}}
\newcommand{\HeI}{He~{\sc i}}
\newcommand{\Hd}{{H$_\delta$}}
\slugcomment{Accepted for ApJ. Tentatively scheduled for Dec. 20, 2000}
\shorttitle{ABUNDANCES IN M33 EARLY B SUPERGIANTS}
\shortauthors{MONTEVERDE, HERRERO, \& LENNON}

\begin{document}

\title{DIFFERENTIAL O AND Si ABUNDANCES IN M33 EARLY B SUPERGIANTS}

\author{M. I. Monteverde\altaffilmark{1}, A. Herrero\altaffilmark{1,2},
\and D. J. Lennon\altaffilmark{3}}
\altaffiltext{1}{Instituto de Astrof\'\i sica de Canarias, E-38200 La Laguna, 
Tenerife, Spain}
 \altaffiltext{2}{Departamento de Astrof\'\i sica, Universidad de La Laguna,
Avda. Astrof\'\i sico Francisco S\'anchez, s/n, E-38071 La Laguna, Spain}
\altaffiltext{3}{Isaac Newton Group of Telescopes, Apartado de Correos 368, E-38700 Santa Cruz de La Palma, Spain}

\date{Received date; accepted date}

\begin{abstract}
We present non-LTE analyses of four M33 early B-supergiant stars and 
five Galactic counterparts. This is the first time that B supergiants 
beyond the Magellanic Clouds are analyzed by means of detailed Non-LTE 
techniques. Among the M33 stars, new spectroscopic observations 
of B38 (ob21--108) are presented and the object is classified
as B1Ia. The classification of another M33 star, B133, is changed with
respect to a former study. Equivalent widths of O and Si lines 
are measured for the M33 objects.
Stellar temperatures, gravities, microturbulences and 
Si abundances are derived for
all objects using the Si ionization equilibrium and the
Balmer line wings. O abundances are then also derived.
Important approximations made during the calculations
are described, and their influence on the results is analyzed
(namely, we set the Lyman resonance lines in detailed balance
during the calculation of the atmospheric structure for
stars cooler than 20\,000 K, and set the \SiIII\, resonance
lines in detailed balance during the line formation calculations
for all models). It is found that these approximations have
no significant effect on the results at any microturbulence.
We found a difference in the derived temperatures of the earlier
Galactic stars as compared to those obtained by 
other authors, which we attribute to the different lines
used for their derivation. A difference can also be
present in the results when using the \SiII/\SiIII\, and the \SiIII/\SiIV\,
ionization equilibria. We conclude that a strict differential analysis 
is needed to detect abundance differences. Thus we compare
results line by line in M33 and
Galactic stars of stellar parameters as similar as possible.
Three of the four M33 stars turned out to be O deficient as
compared to their Galactic counterparts, and only one,
close to the center of M33  (M33 1054) is found to be moderately
O enriched.
From these differential analyses we find that our data
are compatible with a radial O gradient in M33 as that derived
from \HII\, region data: we obtain --0.19$\pm$0.13 or 
--0.20$\pm$0.07 dex kpc$^{-1}$, depending on whether B133 is included
or not. Our data are also consistent with other possibilities such as a
steep increase of the O abundance in the inner region (at projected
distances less than 9 arcmin from the center of M33), followed
by a flat O abundance profile towards the outer parts of M33.
Si shows the same pattern, and it is shown that Si and O correlate
well, as expected for $\alpha$-elements, supporting then the
high value of the O abundance gradient in M33 as compared to the
Milky Way and other nearby spiral galaxies. The results are compared
with those of a more approximate technique, and it is concluded
that this last can be used, attention being drawn to certain problems
that are indicated. As an important additional
point, it is shown that M33 1054 is most probably a single object,
in spite of the bright absolute magnitude found in the literature.   
\end{abstract}


\section{Introduction}

Extragalactic stellar astrophysics, the study of individual stars
in other galaxies, has been getting a strong impulse in recent
years (see Massey \cite{mass98} and Kudritzki \cite{kud98} 
for recent reviews on the subject). 
The extension of spectroscopic techniques appropriate for
stars in our Galaxy to those in other galaxies (in particular, beyond the
Magellanic Clouds) is difficult not only because
of the faintness of the stars and the related difficulty of the
observations, but also because of the lack of
adequate stellar classification schemes (see Lennon \cite{len97};
Monteverde et al. \cite{mont96}, hereafter Paper I) 
and the fact that we will tend to
pick up the brightests and most extreme objects, for which
we lack adequate Galactic comparison standards.

In spite of all these difficulties this study is of great interest
for a number of astrophysical problems. Derivation of individual
stellar abundances in spiral galaxies will allow us to
get radial chemical abundances of a large number of elements
present in the spectra of blue supergiants, from alpha
to iron-group and s-process elements, therefore giving us
invaluable information about the chemical evolution of the galaxy.
The wind momentum--luminosity relationship (WLR,
see Puls et al. \cite{puls96} or Kudritzki et al. \cite{kud99}),
a purely spectroscopic technique that can allow us to derive
extragalactic distances, can be calibrated against 
metallicity using stars in other galaxies, especially in spirals
where we can find different metallicities in the same system,
avoiding uncertainties in the distance moduli. Finally,
stellar evolution can be
studied in different environments, in particular of different
metallicities, where we can compare the number of stars
in a given evolutionary status and their atmospheric properties
and chemical abundances, thus helping us to disentangle the
role of rotation and metallicity in internal mixing processes (see, for
example, Maeder \& Meynet \cite{maemey00};
Maeder \cite{mae99}; Maeder \& Zahn \cite{maez98}; 
Maeder et al. \cite{maeal99}; Venn \cite{venn99};
Heger \cite{heger98}; Herrero et al. \cite{h99}).

All these studies require observations of faint stars with 
large telescopes, as we did in Paper I, where details concerning
the observations and classification problems can be found.
They also require time-consuming model atmosphere calculations
to get the stellar parameters and chemical abundances.
In Monteverde et al. (\cite{mont97}, hereafter Paper II)
we derived the O abundances of B supergiants in M33 and the
Milky Way by first classifying the M33 stars and selecting their
Galactic counterparts among the stars observed by Lennon et al. \cite{len92}
by requiring that they have the same spectral type and luminosity class.
We then adopted  a spectral classification---stellar parameters
calibration from which \teff~ was obtained, and derived gravity,
microturbulence (again in an approximate way) and O abundance.
This approximate technique allowed us to obtain the O radial
abundance gradient for stars in M33. Our results were in good
agreement with those for \HII\, regions (Henry \& Howard \cite{hh95}).
However, this technique had several serious approximations, and
it is important to confirm its applicability if we want to extend it to 
stars in further galaxies, particularly beyond the Local Group
(with the new 8-m class telescopes, McCarthy et al. \cite{mcC97}
estimate that it will be possible to apply spectroscopic methods
to individual stars at 20 Mpc).

In Section~\ref{secobs} we present the observations. Section~\ref{modcal}
is dedicated to a description of the calculations, and Section~\ref{pardet}
to the derived model parameters and abundances. Section~\ref{comms}
presents a discussion of problems and aspects of interest
encountered in the individual analyses, and the discussion of the
results is presented in Section~\ref{disc}. Finally, our conclusions
are given in Section~\ref{concs}.

\section{The observations}
\label{secobs}

The stars used for the present work have been selected among the
objects we have observed until now in different observing runs at La Palma 
using the 4.2-m
William Herschel Telescope and ISIS spectrograph. The spectral 
resolution is about 2 \AA. This resolution is not optimal, but
adequate
for the present analysis, based on strong isolated lines in 
B supergiants, which typically rotate quickly compared with
cooler spectral
types. Each star has been observed in two wavelength 
regions; the blue region
(4000--4600 \AA) contains most of the features needed to classify and
analyze B-type supergiants. The other region is that of H${\alpha}$, 
which is a luminosity-sensitive and also a wind-diagnostic line. 
Details of the observations and spectral reductions, as well
as plots of the stellar spectra,
can be found in Paper I.	 

Spectral classification of M33 stars  is not as easy as 
might be expected because of variations in metallicity
and lack of adequate comparison standards.
Three of the stars analyzed here were classified by 
Monteverde et al. \cite{mont96}, where these problems are discussed,
and were included in our determination of the radial O gradient of
M33 in Paper II. One of the stars used in Paper II has been
excluded here (M33 1354) because it is too cool for the Si
ionization balance to be used as temperature indicator.
We have replaced it with B38, an early B supergiant that we have
classified as B1Ia, following the criteria described in Paper I
(more details are given in \S\    5).
Figure ~\ref{spec} displays the blue spectra of our M33 stars, and
Figure ~\ref{haspec} displays the \Ha\, part of the spectra.

The Galactic standard spectra we use  here have been selected from
the sample by Lennon et al. \cite{len92}. A description of the
observations and reductions, together with a very useful atlas
of Galactic B supergiants can be found there. In Table  ~\ref{lista} we list
the M33 and Galactic stars used for the present paper.



\placetable{lista}

Equivalent widths for the oxygen and silicon lines were measured by 
a non-linear least-squares fitting procedure to fit Gaussian profiles 
to the absorption lines of interest. Blended lines were fitted by 
multiple Gaussians to estimate the equivalent widths of the individual 
components. Data and fits were visually inspected and an interactive 
fit was made to obtain the errors. The estimated errors reflect the range 
of equivalent widths deduced using different local continuum levels and 
the qualitative accuracy of the profile fit. For the Galactic stars
we adopted the equivalent widths from Lennon et al. \cite{len93}
after checking with some cases that our measurements are 
consistent with theirs. In this case, the errors were adapted from the 
qualitative scale given by Lennon et al. \cite{len93}
by assuming an error close to the maximum possible error given in
Lennon et al. \cite{len93}. In a few particular cases we
measured it directly from the spectrum, following the same technique
as for the M33 stars. Typical errors vary between 10$\%$ and
20$\%$, with extreme values at 5$\%$ and 30$\%$.
The measured equivalent widths 
and adopted errors are listed in Table~\ref{eqwidths}.

\placetable{eqwidths}

\section{Model calculations}
\label{modcal}

For the present paper we have calculated a large number of 
model atmospheres, including small coarse grids
around the \teff~ values initially assigned to each spectral type.
We first generated H/He non-LTE model atmospheres using the ALI code 
(Kunze \cite{kunze95}). The models assume
plane--parallel geometry with both hydrostatic and radiative equilibrium but
metal line-blanketing is omitted. We have adopted a solar helium abundance
for all models. Although a few tests we performed showed that the He
abundance has an effect on the Si ionization equilibrium, recent
results seem to suggest that early B supergiants with low rotational
velocities have at most moderate He enrichments (see McErlean et al.
\cite{mc98}; Smith \& Howarth \cite{sh98}) 
and thus we adopted a normal He abundance
(\eps = $N$(He)/($N$(H) + $N$(He)) = 0.09).
This also agrees with the findings of McErlean
et al. \cite{mc99} indicating that spectral analyses of 
Galactic B supergiants are consistent with moderate or zero 
He enrichment.
Non-LTE line-formation calculations were performed using the codes DETAIL and
SURFACE. 
H and He model atoms have been taken from Herrero et al. \cite{h92}.
In the calculation of line profiles of H and He, we occasionally
 had to perform preliminary runs where Lyman lines where set in
detailed balance (see below).
These populations where used in a second run with the Lyman
lines taken fully into account. 
For the calculations of the metal lines, the atomic data used 
were taken
from Becker \& Butler \cite{bb88} for O\,{\sc ii} and 
Becker \& Butler \cite{bb90} for Si\,{\sc ii}/{\sc iii}/{\sc iv}.
Usually the \SiIII~ resonance lines were set into detailed balance.
Microturbulence has been included in the line formation calculations
(i.e., in the statistical equilibrium equations) and in the
formal solutions. In all cases, populations were considered
converged when they reached relative corrections of 10$^{-3}$.    

Two main assumptions have been made
to improve convergence and save computer time whenever possible.
First, below 20\,000 K,
the Lyman series was kept in detailed balance during the calculation
of the model atmospheres. This means that we assume that, due to the
large optical depth, the upwards radiative
rates cancel out the downwards radiative rates exactly, thus allowing
the corresponding terms to be anallytically deleted from the
statistical equilibrium equations. The levels are still connected
through collisions and, indirectly, through the continuum.
As this approximation could be important for
the Si ionization balance used to determine temperatures, we have
studied its effect on the equivalent widths of the relevant Si lines.
Figure~\ref{noly} shows the difference between
the temperature runs of two models calculated at
\teff = 22\,500 K (where the error in setting
the Lyman lines in detailed balance will be greater than 
below 20\,000 K), 
with and without the Lyman
series in detailed balance. The formation depth of relevant lines
of \SiII~ and \SiIII~ are also plotted. We see that the
\SiII~ lines are
formed much deeper than the zone where the changes are relevant.
This is not the case for the \SiIII~ lines, 
which are stronger and their cores form where the temperature
changes begin. However, their equivalent widths vary by less
than 5 m\AA, which is largely below our measurement errors.
The reason is that at the low densities
and high temperatures of the upper layers of early B supergiants,
the local temperature is not the only factor determining the populations,
and thus small changes in the local temperature have a relatively
small influence.

We also show in Figure~\ref{noly2} the populations
of 4s $^3$S$^e$, the lower level of \SiIII\,$\lambda$4553
(which we take as representative of the other lines in the triplet),
where we see
that the changes are only significant in the very upper layers 
of the atmosphere. In addition, the upper and lower levels
of the transition vary in a similar way. Thus the ratio of the
departure coefficients, which determines the line source function, 
remains nearly the same, and the line is only marginally affected.



The second main approximation has been to set the resonance lines of \SiIII~
into detailed balance. We begin by discussing the effect at
a microturbulence velocity of 0 km s$^{-1}$. Again, there are variations in 
the level population which are not large at the depth where the optical
\SiIII~ lines form (see Figure~\ref{nores1}, where
we can again see the
populations of the lower level of \SiIII~$\lambda$4553), and therefore
the changes in the equivalent widths of these lines are small,
of the order of 2--3 m\AA , when the
\SiIII~ resonance lines are set into detailed balance 
in the model with \teff\ = 22\,500K, \g\ = 2.50, \eps\ = 0.09,
independently of the fact that the Lyman lines were or were not set
in detailed balance during the model calculations. In 
Figure~\ref{nores2} we show the ratio of the departure coefficients
of the upper and lower levels (these levels are actually a packing
of individual levels, see Becker \& Butler \cite{bb90}, but this
does not affect our reasoning here) of 
\SiIII\,$\lambda\lambda$4553, 4568, 4574 with
and without the resonance lines of \SiIII~ in detailed balance,
together with the depth where the line core forms. The small
variation of this ratio, which determines the line
source functions, means that only the very center of the line
varies a little, and thus the effect on the equivalent width
is very small (in the example of Figure~\ref{nores2} the
difference in the flux is 0.2$\%$ in the line center,
reaches a maximum of 0.6$\%$ at 0.6 \AA~ from line center
and dissappears at 1 \AA~ from it).
The variations of the
\SiII~ and \SiIV~ lines are of the same insignificant order.

When we repeat these calculations at $\xi$ = 15 km s$^{-1}$ the
variations in the equivalent widths are again of only a few
m\AA~ (thus even reducing the percentage change, because
the equivalent widths are now larger). The reason
is illustrated by Figure~\ref{nores_15}, where we again see
the changes in the population of 4s\,$^3$S$^e$ when the resonance
lines of \SiIII~ are set in detailed balance or not. Comparing
with Figure~\ref{nores1} we see that the changes in the population of
this lower level of \SiIII\,$\lambda$4553 are even lower than for
the case with $\xi$ = 0 km s$^{-1}$. The increase in the number
of absorbing atoms as a consequence of microturbulence makes
the approximation of detailed balance even more realistic, as the
resonance transitions are now even more opaque. Thus the final
equivalent widths are much larger than without microturbulence,
but the differences when setting the resonance lines in detailed balance
or not remain very small (1--3 m\AA ).

In view of the small effects that these approximations introduce,
compared with the measurement errors and taking into
account that we will perform a differential abundance analysis,
we conclude that the approximations made will not have any
influence on our results.




\section{Parameter determination}
\label{pardet}

Our procedure for determining the stellar parameters is as follows.
As Balmer lines are almost unaffected by $\xi$, H$_{\delta}$, and
H$_{\gamma}$  have been used as indicators of log $g$. For each grid value of
\teff~ we have determined a unique log $g$, for which observed and
theoretical Balmer-line profiles  agreed best. 
In this way, preliminary values of (\teff , \g ) were first determined. 
Figure~\ref{fproc1} illustrates
this for the case of M33 1054.
 

For each of these (\teff , \g ) pairs fitting the Balmer lines
we calculated silicon line equivalent widths at different $\xi$. 
A solar silicon abundance is initially assumed. 
For each pair of Si lines of two successive ionization stages, 
a diagram with theoretical equivalent-width ratios is constructed (see 
Figure~\ref{fproc2}). We then compare 
with the observed equivalent width, and interpolate the
values of \teff , \g,~ and $\xi$ that fit this observed
equivalent width. The error bars are used to place limits
to the determined values of the stellar parameters.  This step
of the procedure is illustrated in Figure~\ref{fproc2}, again taken
from the analysis of M33 1054.
One of these plots is produced for each ratio of Si lines
in two successive ionization stages, and the average values of
\teff~ and \g~ are adopted.


We then considered the preliminary value of the (\teff , \g ) pair as fixed
and the microturbulence and Si abundance were freely varied in order to
ensure that the derived abundances from individual lines were not a function
of strength.  Si\,{\sc iii}\, lines, which come from a single multiplet, were
chosen to create the diagram of Si abundance against observed equivalent
width. The new values of microturbulence and Si abundance are obtained by 
demanding a zero slope in the (EW-Abundance) plane of Si\,{\sc iii} lines,
for those abundances compatible with the other ionization stages.

The so defined Si abundance served as a basis for a new (\teff , \g )  
determination. These steps are repeated until the change in
\teff , \g , $\xi$, and Si abundance was within the error in the
equivalent-width measurements. This step is also illustrated by
Figure~\ref{fproc2}. From this figure and those corresponding
to the other line ratios we can read out the parameters and
the error box, which is given by the uncertainty in equivalent
widths. Note here that \teff , \g , $\xi$ and Si abundance are
determined simultaneously, and there is a unique combination
of these four parameters for each point within the error box.
Thus not all combinations of stellar parameters 
are possible and the errors in these
quantities are given by the equivalent widths errors.
Errors have been adopted by taking the largest error box consistent
with all line ratios. Adopted values are $^{+1000}_{-800}$K in \teff , 
$\pm$0.1 in \g~ and $\mp$2 km s$^{-1}$ in $\xi$. For the Si
abundances, we have taken individual abundances for each line and adopted
as error the variation of the abundance within the error box.
Typical error abundance values oscillate between 0.1 and
0.2 dex.

Table~\ref{param} lists the results of our analyses of the
M33 B supergiants and their Milky Way comparison standards.
We explicitly indicate whether the stellar parameters have been
obtained by using the \SiIII/\SiIV\, or \SiII/\SiIII\, equilibria,
since there are some differences in the results, 
as can be seen for example in the derived parameters of M33 1054
or in the O abundances of the Galactic stars.

\placetable{param}

Table~\ref{oaberr} gives the errors in the O abundance
due to the uncertainties in the equivalent width measurements
and in the model parameters for the M33 objects. Note that the situation is
different as for the Si abundance, as now the model parameters
are independent from the derived O abundance. Thus errors have been
added quadratically. To calculate them
we have computed, at different abundances, a fine grid of models around those
given in Table~\ref{param} as defined by the respective error boxes.
We see that final errors are mostly dominated by the uncertainty 
in equivalent widths. The same procedure has been used for
the Galactic comparison stars.

\placetable{oaberr}

\placetable{siab}

\placetable{oab}

Tables~\ref{siab} and \ref{oab} give the Si and O abundances derived for
each individual line used in the study. 
The average values have been calculated by 
weighting individual values with the inverse square errors and
final errors have been calculated using the
quadratic mean of individual errors. We see from the table that
individual lines can give quite different values, but no 
systematic pattern can be appreciated, especially when we consider
the behavior of the Galactic standards. This points to the
comparatively low S/N ratio of the M33 stars as the main reason
for individual variations in the extragalactic sample. 

Absolute abundances are listed, but given the present state of the art of 
B--supergiant model atmospheres, we think that the differential abundances
are more trustwhorthy, and thus we prefer to derive stellar differential
abundances from individual differences, following the ideas presented
by Smartt et al. \cite{sm96}. In Tables~\ref{difsiab} 
and~\ref{difoab} we give the differential Si and O abundance between
the M33 stars and their Galactic counterparts
for each individual line. Again, some individual variations are
very large, but we preferred to retain them all in this work because
of the small number of available lines. The final adopted
differential abundances are given later in \S~\ref{disc}. We have
retained the individual values for each comparison star.

\placetable{difsiab}

\placetable{difoab}

\section{Comments on the spectral analyses}
\label{comms}

In this section we comment on the individual analyses and the
problems found in them. We first consider the analysis of the
Galactic stars as a whole, and later we comment on the M33 stars.

\paragraph{Galactic standards} The Galactic objects we have analyzed here
have been analyzed before by McErlean et al. \cite{mc99} using
similar methods. The first thing that is evident when we compare our
results with those of McErlean et al. \cite{mc99} is a
difference in temperature for some of the stars in common.
In Figure~\ref{scale} we have plotted the temperature scale 
fitted to the data given by McErlean et al. \cite{mc99}
by assigning a number to each spectral type, as well as the
values obtained by McErlean et al. \cite{mc99} and
ourselves for the stars analyzed here. We see that, for B1.5 and above,
the results differ, especially for HD\,13\,854 and HD\,14\,956,
our temperatures being lower by up to 2\,000 K (10 $\%$). Correspondingly,
our gravities are lower and our microturbulence velocities
higher.

We attribute these difference to our use of the \SiIII~ triplet close to
4550 \AA , instead of the weaker \SiIII~ triplet at 
$\lambda\lambda$4813, 4819, 4829 that was chosen by McErlean et al.
\cite{mc99} for their analyses. Of course, our choice is
primarily determined by the observations of the extragalactic
stars, which, because of observing-time limitations, had to
be wavelength restricted. 

The main reason why McErlean et al. \cite{mc99} disregarded
the bluer triplet lines was that they do not behave as
expected. It is expected that their strength increases with
decreasing gravity at a fixed temperature, and, although this is the
general trend, for very low gravity models McErlean et al. \cite{mc99}
find a filling-in of these lines that they do not consider to be  real
and attributable to the plane--parallel approximation (which 
leads to unrealistically low densities in the upper layers
of the model atmosphere). The \SiIII\,$\lambda\lambda$4813, 4819, 4829
lines, being weaker, should have a better and more realistic 
behavior in the plane--parallel models. 
We agree with the results of McErlean et al. \cite{mc99}, although
we find a similar, less marked, trend in the 
\SiIII\,$\lambda\lambda$4813, 4819, 4829 lines.

Given the tests described in the preceding section, we do not
consider  the approximations made there to be the reason
for the temperature differences, but rather that it is to be attributed to the
use of two different criteria. Of course, with the limitations
of the plane--parallel, hydrostatic models the weaker lines should be
preferred, although they are also of more limited application
and for the present work, with very faint extragalactic stars,
these lines would have had large errors in the equivalent
width measurements. The question of which particular criteria are to be 
preferred can be completely solved only with the use of
spherical models with mass-loss, which are more adequate for early
B supergiants. Of course, we are implicitly assuming 
that with a differential analysis such as the one presented here the 
influence of the particular criteria adopted will only be of
secondary importance.

We can also see in Table~\ref{oab} that the Galactic stars show a
trend towards larger O abundances for those stars with parameters determined
through the \SiIII/\SiIV\, ionization equilibrium, as compared
to those for which the \SiII/\SiIII\, equilibrium was used.
The difference is large (0.4 dex on average) and the 
first idea would be to consider that this is related to the
above discussion concerning the differences in temperature with McErlean
et al. \cite{mc99}. Note, however, that
the stars for which we obtain the
temperature differences with McErlean et al. \cite{mc99} are just those 
for which the derived O abundance is solar, or close to it, in
agreement with what might be expected.  This again justifies
that comparisons be made in a strictly differential way.

Another peculiarity is the high Si abundance found for HD\,14\,956,
a factor of two larger than the solar abundance. This comes mainly
from the low temperature and gravity, which would mean that this
star is an extreme supergiant. This is supported by its luminosity class
and the strong P Cygni profile displayed in \Ha\, (slightly stronger than that
displayed by HD\,13\,854 which is hotter, but has a larger gravity and
lower luminosity class). Due to this particularity of HD\,14\,956 we will
always use a second Galactic supergiant when deriving differential
abundances for the M33 stars.
 

\paragraph{M33 1054} The analysis of M33 1054 is particularly
difficult, because of the possibility of a multiple spectrum
and the weakness of the \SiIV~ and \SiII~ lines. 
In HST WFPC2 images taken with the F439W filter
(cycle 6, P.I. J.K. Mccarthy) we can see a cluster of stars
at the position of M33 1054. This cluster is dominated by one
object that corresponds to the bright star we observed.
However,
there are two
indirect indications of multiplicity. Firstly, the absolute
magnitude of the star is very bright for a B2Ib star. 
Ivanov et al. \cite{iv93}
give $m_V$ = 16.6, which, combined with a mean reddening
of $E(B-V)$ = 0.16 (Massey et al. \cite{mas95}), a true distance
modulus of 24.64 (Freedman et al. \cite{free91}), and an $R$ value
of 3.1, gives an absolute visual magnitude of $M_V$ = --8.5. Even using 
the lowest value found by Massey et al. \cite{mas95} for
the M33 reddening, we find $M_V$ = --8.3, which is extreme
for a B2Ib supergiant, although in view of the large scatter
present in the spectral type--absolute magnitude calibrations
we do not consider this as a definitive proof.

Secondly, the identification of the
\SiII~ doublet is dubious. The observed positions of the \SiII~ doublet lines
$\lambda\lambda$4128, 4131 are 4128.88 and 4132.06 as determined from
a double-Gaussian fit. There are no other lines to which these positions
could be clearly assigned. This means that the lines seem to be 
displaced with respect to the laboratory position by 0.81 and
1.17 \AA, respectively. This could be interpreted as due to a composite
spectrum, especially taking into account the position
of M33 1054 (its projected distance to the center of
M33 is only 1.0 arcmin). 
However, we cannot see any other sign of multiplicity in the blue
spectrum. In particular, the \SiIII~ lines do not show indications
of a secondary spectrum (nor does the \SiIV\,$\lambda$4116 line).
This means that any possible contamination should come from a
spectral type with strong \SiII~ lines and weak \SiIII~ ones,
i.e., around A0. We have tried to attribute the \SiII~ lines to such
a possible cooler companion and have made some numerical simulations
using the Galactic standards from Lennon et al. \cite{len92}.
We have taken the spectrum of HD\,14\,899, an A0Ib star, 
and have combined it with that of HD\,206\,165, a B2Ib star
(we have also used HD\,21\,291, a B9Ia star, to be sure that
the luminosity class would not influence the conclusions).
After displacing the spectrum of the A0Ib star,
the spectra of the A0 and B2 supergiants were combined in a 1:3 ratio 
for the blue, and 
1:1.6 for the red (the ratios were derived assuming that
the observed \SiII\, lines are only due to the possible A0 
companion; they correspond to a ratio
of the stellar radii of 2:1). We then degraded the resolution
and the S/N ratio of the composite spectrum to those of
the observed spectrum of M33 1054 and compared them.
The same procedure was followed with the spectrum of
HD\,206\,165 separately.
In Figure~\ref{bin_blue} we compare the simulations in the blue
and in Figure~\ref{bin_red} in the red.
In the blue (Figure~\ref{bin_blue}) we see that the apparent displacement
of the \SiII\, $\lambda\lambda$4128, 4130 lines
can be reproduced from the the spectra of the single blue star,
while the \SiII\, lines should have been more conspicuous were 
they to be attributed
to a composite spectrum with the given ratio. 
In Figure~\ref{bin_red} we see that the \SiII\,$\lambda\lambda$6347, 6471 
lines should be
apparent, although the spectrum is marginally compatible with
the observations due to our low S/N ratio in the red.
Here, as in the whole spectrum, the qualitative agreement between
the simulations and the observed M33 1054 spectrum is better for the
single B2Ib spectrum, although we cannot completely rule out
the presence of the A0 star. For example, we also see in the
red that the \CII\,$\lambda$6578.03 line is too strong in
the simulated single spectrum.
Of course, the spectral ratios could be lower
than assumed, but then the blue star dominates the observed
spectrum more and more. However, as the degradation of the S/N was made by generating 
a random array of normally distributed numbers with an appropriate 
standard deviation, the simulations are not unique results, and thus
we must be careful with our conclusions.

We should also point out that the possible
presence of an A0Ib star does not improve the situation
with respect to the bright magnitude, because it would
result in a combined magnitude only 0.3 magnitudes brighter
than that of the B supergiant. 
In addition, the equivalent-width ratio 
\SiII(4128+4131) / \SiIII(4553+4568+4574) is similar to that
of HD\,206\,165, a B2Ib Galactic supergiant, while
the  \SiIV(4116) / \SiIII(4553+4568+4574) ratio is only slightly
lower than that of HD\,13\,841, a Galactic B1.5Ib star.
An additional indication that M33 1054 is dominated by
a single object comes from
the work of Massey et al. \cite{mass96}. 1054 can be identified
with their UV source UIT 215, for which they find a spectral
type of B3I, but no signs of multiplicity in the optical spectrum,
the UV photometry, or the {\it HST\/} images.
Thus the observed spectrum of M33 1054 is perfectly compatible
with a star slightly hotter than HD\,206\,165 (B2Ib) and
slightly cooler than HD\,13\,841 (B1.5Ib), except for
the displacement of the \SiII~ lines at $\lambda\lambda$4128, 4131.



Thus if we assume our spectrum of
M33 1054 to be dominated by a single object, we can reproduce
in many simulations the apparent displacement of the \SiII~ lines,
while if we assume it
to be at least binary, 
we would expect the \SiII\, lines to be more conspicuous,
or have to give the cool companion a very low surface brightness,
thus minimizing its contribution to the combined spectrum.

For all these reasons, we concluded that the analysis of
M33 1054 would not be seriously compromised by the presence
of a possible companion. However, for safety,
we have decided to analyze M33 1054 using 
the \SiII~ and \SiIV~ lines separately, once comparing with
HD\,206\,265 and using the \SiII/\SiIII~ ionization
balance, and a second time comparing with HD\,13\,841
and using the \SiIII/\SiIV~ one. We see in Table~\ref{param}
that the results are
quite similar, not only with regard to the stellar parameters
\teff, \g, and $\xi$, but also with regard to the 
abundances obtained. This gives us strong confidence in our
conclusion that M33 1054 is a slightly metal-enriched blue supergiant,
a result similar to that found for this star in Paper II.

The fit to the observed M33 1054 spectrum can be seen in Figure~\ref{1054fit},
where we have plotted the H, He, Si, and O lines in the given wavelength 
interval. We show here only the fit with the parameters from the
\SiII/\SiIII\, ionization equilibrium (that with the \SiIII/\SiIV\,
one is comparable).
We see that the overall agreement is excellent, except for
the \OII\,$\lambda\lambda$ 4276, 4415, 4417 lines. The first one
is affected by the level setting of the local continuum, and the last two
are blended with the
strong broad interstellar absorption band at 4428 \AA . \HeI\, lines
generally fit  well and do not demand an increase in the helium
abundance, although such an effect would be hard to see with the
quality of the M33 1054 spectrum.


\paragraph{M33 B133} This is an important object, because it defines the
central part of our radial M33
gradient (see next section). In Paper I,
we classified the star as B2.5Ia. When 
determining its \teff~ during
this work we realized that the temperatures obtained
were much larger than those corresponding to a B2.5 supergiant.
The star is more of a B1.5Ia star (we acknowledge P. Massey here, who already
indicated  this possibility to us after publication of our Paper I).
As a consequence (and also as a cause) we classified the star
as a normal-metallicity object, which again evidences the 
interplay between metallicity and stellar classification.
The fact that we found a strong metal deficiency in this
star in our analysis of Paper II was already an indication 
of the problem. Thus we here correct our previous classification, 
as given above in Table~\ref{lista} and compare with stars
of similar temperature (this is the reason why we analyzed HD\,198\,478
at the beginning, but did not use it in the final comparisons; 
nevertheless, we decided to keep it in the tables for future reference).

We first compared it with HD\,13\,854, a B1Iab star. While
we obtained similar values for the stellar parameters
\teff, \g, and $\xi$ (see Table~\ref{param}) the O and
Si abundances derived were widely different. M33 B133
was seen to be clearly deficient in O and Si with respect to
HD\,13\,854. The large deficiency lead us to use
HD\,13\,841 (B1.5Ib) also as comparison standard, with
comparable results, as can be seen in Table~\ref{param}.
Thus we conclude that M33 B133 is largely deficient in
$\alpha$--elements when compared with stars in the
solar neighborhood.

This result is qualitatively similar to that obtained
in Paper II, where we compared M33 B133 with the Galactic 
B supergiant HD\,198\,478, a B2.5Ia star. However, the differential
abundance of O obtained in Paper II (--0.45 dex) was more moderate than that
 obtained here (--0.65 dex). This is an indication that
differential misclassification of the stars can introduce important
errors in the differential analysis, and that this is the
main disadvantage of the technique used in Paper II.

\paragraph{B38 (ob21--108)} This star was not included in our previous
papers, but we have included it here because it is an early B-supergiant
far away from the center of M33, and thus an appropriate substitute of 
M33 1354, which could not be analyzed here because the Si
ionization balance cannot be used as temperature indicator.
It has been classified as B1.5Ia$^+$ by Massey et
al. \cite{mas95}. The absence of the \SiII~ doublet at 
$\lambda\lambda$4128, 4130 \AA~ indicates a
spectral type earlier than B2. The line ratios of \SiIV\,$\lambda$4088 and
the \OII~ features around it point to a B1--B1.5 spectral type. The
presence and strength of \SiIV\,$\lambda$4116  suggests 
that a B1 spectral type
would be more appropriate. Finally, a comparison of 
Mg\,{\sc ii} $\lambda$4481 and the
\SiIII\, triplet around 4550 \AA~ agrees well with a B1 classification. The red
spectrogram shows an emission in \Ha, which along with the relatively
narrow H wings, is consistent with a Ia luminosity class. Our spectral
classification for ob21-108, then, is B1Ia, which is consistent
with Massey et al. \cite{mas95}.

The spectral analysis of this star resulted in a lower temperature
than in the case of B133 and was comparable to that of 1054 in spite
of the earlier spectral type. This is a consequence of a 
lower gravity and 
larger microturbulence. The abundances, however, are intermediate
between those of 1054 and B133, pointing to a slightly metal-deficient star.

\paragraph{110--A} 
The blue spectrum displayed in Figure~\ref{spec} is actually
a merger of spectra obtained in two different observational
runs (described in Herrero et al. \cite{h94} and in Paper I). 
We first checked the overlapping parts for agreement,
as we have used the \SiIII\, lines from the first observation
with the \SiIV\, lines of the second  to derive the stellar
parameters from the Si ionization balance, as described in the 
preceding section. We could not find any variation in the blue spectrum.

This object has a very strong \Ha\, emission, indicating
the presence of a strong wind (see Figure~\ref{haspec}). It is clearly 
the most extreme object in our small extragalactic sample, 
with a very low gravity and effective
temperature, this last in spite of the relatively early spectral type.
This can also be seen in the \Hg\, profile, filled in with emission
(see Figure~\ref{spec}). For this reason, only \Hd\, (particularly the wings)
was used for the gravity determination. Correspondingly, the derived 
microturbulence is again very large. The results of the last two
objects, 110-A and B38, seem to suggest that the derived microturbulences
could in part be due to the presence of strong winds, although this
is not reflected by the values obtained for the Galactic stars
(see Table~\ref{param}) and actually cannot be stated before a complete
wind analysis is done, which is beyond the scope of this paper.
  
\section{Discussion}
\label{disc}
In this section we would like to discuss two points: the O and Si
abundance gradients in M33 derived from the differential analysis
of the Galactic and M33 stars, and the comparison of the results
obtained in this paper with those obtained with the
more approximate (but much less time-consuming) technique used in
Paper II. 

\paragraph{The O and Si gradients in M33} 
Table~\ref{grads} gives the differential
Si and O abundances of M33 stars with respect to their Galactic counterparts
(obtained from the data in Tables~\ref{difsiab} and \ref{difoab}),
as well as the projected distance of each to the center of M33
(in arcmin). Average values and errors have been obtained in
the same way as in Section~\ref{pardet}.

\placetable{grads}

In Figure~\ref{ograd} we have represented the differential O abundances of the
M33 stars versus their projected distance from the center of M33. We
have retained both values from Table~\ref{grads} in the plot to
get a feel for the present uncertainty level. We also plot in 
Figure~\ref{ograd} the O gradient derived by Henry \& Howard \cite{hh95}
from \HII\, regions data (actually, from data obtained by several authors).
We see that only B133 departs significantly from the gradient
given by Henry \& Howard \cite{hh95}. One could conclude 
(doing some low-number statistics!) that the
stellar data are not inconsistent with the results from \HII\, regions
and that although more data are needed, as usually claimed in astrophysics,
a first guess could be that B133 is a comparatively low metallicity
star for its radial position in M33 (of course, one could also
find more exotic explanations). 
In fact, from the three objects that agree with Henry \& Howard \cite{hh95}
we derive a gradient for the O abundance
in M33 of --0.20$\pm$0.07 dex kpc$^{-1}$, which
 coincides with that of Henry \& Howard (--0.22$\pm$0.04 dex kpc$^{-1}$).
Also, including M33 B133 we obtain the same value, with a larger error
(--0.19$\pm$0.13 dex kpc$^{-1}$). However, without the visual guide
provided by the solid line in Figure~\ref{ograd} one would
not immediately conclude that there exists a linear gradient in
the stellar abundances of the M33 stars, but merely that the star
near the center (1054) is much more O rich that those beyond 9 arcmin
(about 1.8 kpc at the distance of M33)
from the center. Thus, a step function would be equally good for
representing the stellar data.

This, of course, is due to our lack of data, but indicates
that there are other possibilities, apart from a linear radial gradient.
The gradient derived from Henry \& Howard \cite{hh95} is the final
result obtained after combining results from different groups.
In Figure~\ref{ohflat} we have represented our values
(the average values of Table~\ref{grads} now for clarity) 
together with those
of Vilchez et al. \cite{pepe88}, one of the groups whose data
were used by Henry \& Howard \cite{hh95}. The data of Vilchez et al.
\cite{pepe88} are represented relative to the solar value (while
our values are relative to Galactic standards).
We see there that
a steep gradient in the inner regions, combined with a flattening
(or even a constant value) beyond 10 arcmin would be also
consistent with this particular data combination. Thus
we simply conclude that the few stellar data
are consistent with any possibility of a larger O abundance
in the inner regions of M33 than in regions at around 2 kpc (projected
distance) and beyond. Obviously, we need to extend our database.



Figure~\ref{sigrad} shows the Si differential abundances in M33 from the 
data in Table~\ref{grads}. This time we see that the differences between
the two individual values corresponding to the same object are
larger for the outer stars, reflecting the largely different Si abundance
derived for the Galactic stars HD\,14\,956 and HD\,13\,854. 
In spite of this problem, we
see that the trend is the same as for the O data, namely that the
star at the center has a much larger Si abundance than the outer objects.
This similarity is corroborated in Figure~\ref{siocorr}, where
we have represented the Si and O differential abundances found for the
M33 stars. It is obvious that they
correlate very well, indicating a common origin,
in agreement with our present understanding of chemical 
stellar and galactic evolution(see for example McWilliam, \cite{mcw97}).

The good correlation between the O and Si gradients has an added
value, as it confirms the large O abundance gradient of M33, much larger
than that of the Milky Way ($-$0.07 dex kpc$^{-1}$,
see Gummersbach et al. \cite{gumm98} for a list of values), 
M31 ($-$0.04 dex kpc$^{-1}$, see Blair et al. \cite{blair81})
or M81 and M101 (with values of $-$0.06 and $-$0.11 dex kpc$^{-1}$
respectively, see Henry and Howard \cite{hh95}). If finally confirmed
(as we still have a very limited number of data points) this points to very
interesting questions regarding the physical reason for this
difference, and would set important constrains to the galatic
chemical evolutionary models. 



\paragraph{Comparison with the technique in Paper II}
In Paper II we used an approximate technique to derive the stellar
M33 O gradient, whose results we want to compare with the
more detailed analysis carried out here. Although all the details
of the technique are explained in Paper II, we give here a short
description, for greater ease and better understanding of the
differences found.

In Paper II
we derived the O abundances of B supergiants in M33 and the
Milky Way by first classifying the M33 stars and selecting their
Galactic counterparts among the stars observed by Lennon et al. \cite{len92}
by requiring that they have the same spectral type and luminosity class.
We then adopted a spectral classification--stellar parameters
calibration from which \teff~ was obtained.
The gravity was then 
derived by fitting the wings of \Hd~ at this fixed temperature.
We then assumed
that the Galactic B supergiants have solar abundances, and then
derived the microturbulence of each star in the Milky Way. This
microturbulence was then adopted for the corresponding M33 star,
and the O abundance then followed directly by comparing the observed
equivalent width of \OII~4366 with the appropriate line-formation
calculations. This procedure has several important drawbacks: a) there
could be a {\it differential} misclassification in one of the
stars; b) there could be a difference in the stellar microturbulence
of the M33 and the Galactic star; c) the O abundance of the Galactic 
star could be different from solar; and d) only one line of \OII~ was
used. Each of these points could lead to relative differences
in the derived O abundances, introducing systematic errors in the
derived radial gradient of M33.

In Figure~\ref{orelat} we have represented the differential O abundance
obtained in this work versus the same quantity obtained in 
Paper II for the three stars in common (1054, B133, and 110--A).
We see from the figure that the agreement is good,  in spite of the fact
that all four possibilities mentioned above have occurred at least once.


The effect of the use of more than one line was investigated
by Monteverde \cite{mont98}, who used the technique from Paper II
with two lines instead of one, obtaining differences with the
results of Paper II quite similar to those obtained here
for 1054 and 110-A. Also, a slightly larger relative abundance than in 
Paper II was obtained by Monteverde \cite{mont98} for B133.
Thus, not only does B133 have 
the largest deviation from the 1:1 line in Figure~\ref{orelat}, but also
the particular correction due to the wrong spectral classification
is larger than that suggested by the distance from the 1:1 line
in Figure~\ref{orelat}. We conclude that
this is the main inconvenience of the technique used in Paper II,
although even this results in corrections that do not change the qualitative
picture, especially if a sufficient number of objects can be analyzed
at a sufficient number of positions.

\section{Conclusions}
\label{concs}

We have presented non-LTE analyses of four M33 early B-supergiant stars and 
five Galactic counterparts. The first important conclusion of our analyses
is that normal, standard quantitative spectroscopic analyses of
blue supergiants in galaxies beyond the Magellanic Clouds are
perfectly possible even with 4-m class telescopes. Although,
of course, a better resolution and S/N ratio than those used here
are highly desirable and easily attainable with 8-m class telescopes.

New spectroscopic
observations of B38 (ob21--108) have been presented 
and the object is classified as B1Ia, in good agreement
with previous classifications in the literature. The 
former classification of B133 (B2.5Ia) has been changed
to B1.5Ia. We have shown that M33 1054 is most probably a single object,
in spite of the large absolute magnitude found in the literature,
very unusual for a B2Ib star.

We have shown that it is possible to set the Ly lines in detailed balance
during the calculation of the atmospheric structure without affecting the
results, in so far as these are based on photospheric lines formed
sufficiently deep in the photosphere. The same conclusion has been
reached when setting the \SiIII\, resonance lines into detailed balance
during the line-formation calculations, i.e., during the joint
resolution of radiative-transfer and statistical-equilibrium
equations. We have shown that this last approximation is even better
suited when a large microturbulent velocity is present.

We found a difference in the derived temperatures of the earlier
Galactic stars as compared to those obtained by McErlean et al. 
\cite{mc99} which we attribute to the different lines
used for the temperature derivation. Together with some other facts,
such as the apparent correlation
of O abundance with the particular Si ionization equilibrium used,
we conclude that more work is needed before we can trust the
absolute abundances derived, and that strict differential analyses
are required to perform abundance comparisons.

We  then performed a strict differential analysis 
by comparing results line by line in M33 and
Galactic stars with stellar parameters as similar as possible.
Two standards were found for each M33 star, even if
this resulted in comparing stars differing a
little in their stellar parameters to avoid 
a particular result  having a strong unbalancing influence.
From this differential analysis we find that our data
are compatible with a radial O gradient in M33, such as that derived
from \HII\, regions data by Henry \& Howard \cite{hh95},
although one of the four stars departs
from the relation, but also with other possibilities, such as a
steep increase in the O abundance in the inner region (at projected
distances of less than 9 arcmin from the center of M33), followed
by a flat O abundance profile towards the outer parts of M33.
Si shows the same pattern, and it is shown that Si and O correlate
well, as expected for $\alpha$-elements. 

Clearly, more and better data are needed to take this work further,
covering one radial direction in M33 with sufficient spatial resolution.
Note that at present all objects at the same projected radial distance
from the center in any direction are presented together, 
in spite of the clear inconveniencies of this approximation,
because it will immediately add a non-physical scatter to the results.
Also, the error bars should be reduced, and for this better observations
allowing the use of more lines in the analysis are needed.
 
The comparison of results obtained here 
with those of the more approximate technique presented in Paper II
suggests that this last can be used if a careful spectral classification
at the adequate metallicity is performed.

\acknowledgements{We warmly acknowledge Stephen Smartt for 
many ideas, discussions and
suggestions during the realization of this work. Neil McErlean and 
Phil Massey are also acknowledged for their contributions through numerous
discussions. Terry Mahoney is acknowledged for carefully reading and
correcting the manuscript. Our anonymous referee is acknowledged for
her/his criticisms, that improved the scientific value of this paper.
AH acknowledges
support for this work from the Spanish DGES under project PB97-1438-C02-01
and from the Gobierno Auton\'omico de Canarias under project
PI1999/008. The WHT is operated on the island of 
La Palma by the RGO in the Spanish Obervatorio del Roque de los Muchachos 
of the Instituto de Astrof\'\i sica de Canarias.}

\clearpage

\input{table1}

\clearpage

\input{table2}

\clearpage

\input{table3}

\clearpage

\input{table4}

\clearpage

\input{table5}

\clearpage

\input{table6}

\clearpage

\input{table7}

\clearpage

\input{table8}

\clearpage

\input{table9}

%
\clearpage
\include{figus}
\end{document}

%% file: table1.tex
\begin{deluxetable}{llllll}
\tablecolumns{6}
\tablewidth{12.5cm}
\tablecaption{M33 Program Stars and Galactic Counterparts. \label{lista}}
\tablehead{
\multicolumn{3}{c}{M33 stars} & \colhead{} & 
\multicolumn{2}{c}{Galactic stars} \\
\cline{1-3} \cline{5-6} \\
\multicolumn{2}{c}{Ident.} & \colhead{Spec. type} & \colhead{} &
\colhead{HD number} & \colhead{Spec. type} \\
}
\startdata
110A  & 0785       &  ~~B1 Ia$^{+}$     & & HD13854  & ~~B1Iab\\
B38   & ob21--108  &  ~~B1 Ia           & & HD14956  & ~~B1.5Ia    \\
B133  & 0515       &  ~~B1.5 Ia         & & HD13841  & ~~B1.5Ib \\
      &	1054       &  ~~B2 Ib 		& & HD206165 & ~~B2Ib  \\
      &            & 			& & HD198478 & ~~B2.5Ia \\
\enddata
\tablecomments{M33 stars spectral classification is from Paper I. The 
identification numbers are from Humphreys \& Sandage (\cite{hs80}, 
first column) and Ivanov et al. (\cite{iv93}, second column). 
Compared with Paper I, B38 is a new object and B133 has been reclassified.
The Galactic comparison stars from Lennon et al. (\cite{len92}) are also
listed.}
\end{deluxetable}


%% file: table2.tex
\begin{deluxetable}{lcccccccccc}
\tablecolumns{11}
\tabletypesize{\scriptsize}
\tablecaption{Equivalent widths of the O and Si lines in M33 stars and Galactic
standards. \label{eqwidths}}
\tablewidth{0pt}
\tablehead{
\colhead{M33 Star } & \colhead{Si\,{\sc iii}}   & \colhead{Si\,{\sc iii} }   &
\colhead{Si\,{\sc iii} } & \colhead{Si\,{\sc iv}} & \colhead{Si\,{\sc ii}}   &
\colhead{Si\,{\sc ii}}  & \colhead{O\,{\sc ii}} & \colhead{O\,{\sc ii}} &
\colhead{O\,{\sc ii}}   & \colhead{O\,{\sc ii}} \\
\colhead{} & \colhead{4553} & \colhead{4568} & \colhead{4574} &
\colhead{4116} & \colhead{4128} & \colhead{4130} & \colhead{4317} & 
\colhead{4319} & \colhead{4366} & \colhead{4661} \\ 
}
\startdata
0785 -- 110A	 & 420$\pm$30 & 300$\pm$20 & 200$\pm$20 & 85$\pm$15  & 
	\nodata & \nodata & 85$\pm$15  & 150$\pm$22 & 110$\pm$10 & 205$\pm$25 \\
ob21-108 -- B38  & 410$\pm$30 & 370$\pm$20 & 165$\pm$20 & 135$\pm$15 & 
        \nodata & \nodata & 140$\pm$15 & 220$\pm$20 & 155$\pm$15 & 150$\pm$20 \\
0515 -- B133     & 350$\pm$30 & 210$\pm$20 & 165$\pm$20 & 80$\pm$15  & 
        \nodata & \nodata & 65$\pm$20  & 110$\pm$20 & 105$\pm$20 & \nodata \\
1054	 & 300$\pm$20 & 270$\pm$30 & 170$\pm$20 & 65$\pm$10  & 70$\pm$15 
	 & 120$\pm$20 & 120$\pm$10 & 100$\pm$20 & 125$\pm$15 & 200$\pm$30 \\
\tableline
\sidehead{Galactic Star}
\tableline
HD13854  & 430$\pm$30 & 370$\pm$26 & 225$\pm$16 & 160$\pm$11  & 
        \nodata & \nodata & 190$\pm$13 & 210$\pm$15 & 215$\pm$15  & 245$\pm$17 \\
HD14956  & 450$\pm$31 & 405$\pm$28 & 255$\pm$18  & 105$\pm$7  & 
        \nodata & \nodata & 155$\pm$11 & 170$\pm$25  & 135$\pm$9  & 165$\pm$11 \\
HD13841  & 380$\pm$26 & 335$\pm$24 & 205$\pm$14  & 80$\pm$12 & 
        \nodata & \nodata & 145$\pm$11 & 160$\pm$11  & 160$\pm$10 & 205$\pm$15 \\
HD206165 & 335$\pm$23 & 280$\pm$20 & 145$\pm$10  & 35$\pm$10   & 90$\pm$15
	 & 110$\pm$20 & 105$\pm$8 & 105$\pm$15  & 122$\pm$10  & 160$\pm$10\\
HD198478 & 245$\pm$17 & 200$\pm$14 & 135$\pm$9  & \nodata     & 120$\pm$8
	 & 160$\pm$11 &  75$\pm$18 & 60$\pm$15   & 60$\pm$15   & 70$\pm$15 \\
\enddata
\end{deluxetable}


%% file: table3.tex
\begin{deluxetable}{lllccccc}
\tablecaption{Non-LTE Atmospheric Parameters for the Program Stars. \label{param}}
\tablehead{
\multicolumn{2}{c}{M33 star} & \colhead{\teff} & \colhead{\g} & 
\colhead{$\xi$} & \colhead{Ion. balance} & \colhead{$12 + \log({\rm Si/H})$}& 
\colhead{$12 + \log({\rm O/H})$}
}  
\startdata
110A & 0785     &  19000  & 2.20  & 19 & \mbox{Si\,{\sc iii}}/\mbox{Si\,{\sc iv}}& 7.32$\pm$0.15 & 8.55$\pm$0.17 \\
B38  & ob21-108 &  20700  & 2.37  & 20 & \mbox{Si\,{\sc iii}}/\mbox{Si\,{\sc iv}} & 7.24$\pm$0.12 & 8.55$\pm$0.13 \\
B133 & 0515     &  21300  & 2.53  & 15 & \mbox{Si\,{\sc iii}}/\mbox{Si\,{\sc iv}}  & 7.04$\pm$0.15 & 8.22$\pm$0.20  \\
     & 	1054	&  20500  & 2.50  & 9  & \mbox{Si\,{\sc iii}}/\mbox{Si\,{\sc
 iv}}   & 7.61$\pm$0.18   & 8.90$\pm$0.22 \\
     &  1054    &  21000  & 2.55  & 9  & \mbox{Si\,{\sc ii}}/\mbox{Si\,{\sc
 iii}}  & 7.54$\pm$0.14 & 8.79$\pm$0.20 \\
\tableline
\sidehead{Galactic star}  
\tableline
HD13854 & & 21700 & 2.52 & 14 & \mbox{Si\,{\sc iii}}/\mbox{Si\,{\sc iv}} & 7.59$\pm$0.11 & 9.00$\pm$0.13\\
HD14956 & & 19300 & 2.28 & 15 & \mbox{Si\,{\sc iii}}/\mbox{Si\,{\sc iv}} & 7.91$\pm$0.13 & 8.85$\pm$0.16\\
HD13841 & & 21000 & 2.60 & 15 & \mbox{Si\,{\sc iii}}/\mbox{Si\,{\sc iv}} & 7.47$\pm$0.12  & 8.81$\pm$0.12 \\
HD206165 & &  20000 & 2.45 & 15 & \mbox{Si\,{\sc ii}}/\mbox{Si\,{\sc iii}} & 7.29$\pm$0.09 & 8.59$\pm$0.14\\
HD198478 & & 18500 & 2.20 & 8 & \mbox{Si\,{\sc ii}}/\mbox{Si\,{\sc iii}} & 7.55$\pm$0.09 & 8.41$\pm$0.29 \\
\enddata
\tablecomments{ Atmospheric parameters have been derived from
the silicon ionization balance. Typical errors are 
{\bf $^{+1000}_{-800}$}K for \teff ,
$\pm$ 0.1dex for \g, and $\pm$2 km s$^{-1}$ for $\xi$. The two lines for
1054 give the two results for the different ionization balance used.}
\end{deluxetable}


%% file: table4.tex
\begin{deluxetable}{ccrrrrrrrrrr}
\tablecolumns{12}
\tabletypesize{\scriptsize}
\tablecaption{O abundance errors of the program stars. \label{oaberr}} 
\tablehead{
\colhead{ O\,{\sc ii}} & & \colhead{110A} & \colhead{B38} & \colhead{B133} &
\colhead{1054} & \colhead{1054} & \colhead{HD13854} & \colhead{HD14956} &
\colhead{HD13841} & \colhead{HD206165} & \colhead{HD198478} \\
line & & & & & \colhead{cool} & \colhead{hot} & & & & & 
}
\startdata
4317 & model A    &  0.08 &  0.03 &  0.06 &  0.06 &  0.05 &  0.05 &  0.08 &  0.08 &  0.09 &  0.05 \\  
     & model B    &  0.10 &  0.06 &  0.07 &  0.08 &  0.09 &  0.07 &  0.08 & -0.02 &  0.09 &  0.11 \\
     & model C    & -0.02 & -0.06 & -0.06 & -0.04 & -0.03 & -0.04 & -0.10 & -0.09 & -0.09 & -0.06 \\
     & model D    & -0.04 & -0.06 & -0.08 & -0.08 & -0.07 & -0.06 & -0.11 & -0.11 & -0.09 & -0.10 \\
     & $\Delta$EW & $\pm$0.15 & $\pm$0.10 & $\pm$0.24 & $\pm$0.15 & $\pm$0.10 & 
                    $\pm$0.10 & $\pm$0.09 & $\pm$0.08 & $\pm$0.07 & $\pm$0.30 \\
     & Adopted    & $\pm$0.17 & $\pm$0.12 & $\pm$0.25 & $\pm$0.17 & $\pm$0.13 &
                    $\pm$0.12 & $\pm$0.13 & $^{+0.11}_{-0.14}$ & $\pm$0.11 & $\pm$0.32 \\

4319 & model A    &  0.09 &  0.03 &  0.06 &  0.06 &  0.04 &  0.05 &  0.09 &  0.07 &  0.09 &  0.06 \\  
     & model B    &  0.10 &  0.06 &  0.07 &  0.09 &  0.08 &  0.06 &  0.08 & -0.01 &  0.09 &  0.10 \\
     & model C    & -0.03 & -0.06 & -0.06 & -0.04 & -0.04 & -0.04 & -0.09 & -0.08 & -0.09 & -0.06 \\
     & model D    & -0.04 & -0.07 & -0.08 & -0.07 & -0.08 & -0.07 & -0.10 & -0.09 & -0.09 & -0.10 \\
     & $\Delta$EW & $\pm$0.16 & $\pm$0.11 & $\pm$0.15 & $\pm$0.25 & $\pm$0.20 & 
                    $\pm$0.10 & $\pm$0.20 & $\pm$0.08 & $\pm$0.14 & $\pm$0.26 \\
     & Adopted    & $^{+0.19}_{-0.16}$ & $\pm$0.13 & $\pm$0.17 & $\pm$0.27 & $\pm$0.22 &
		    $\pm$0.12 & $\pm$0.22 & $\pm$0.12 & $\pm$0.17 & $\pm$0.28\\

4366 & model A    &  0.08 &  0.03 &  0.02 &  0.03 &  0.01 &  0.03 &  0.06 &  0.05 &  0.05 &  0.04 \\  
     & model B    &  0.10 &  0.07 &  0.03 &  0.08 &  0.07 &  0.06 &  0.07 & -0.01 &  0.06 &  0.11 \\
     & model C    & -0.02 & -0.05 & -0.09 & -0.03 & -0.03 & -0.04 & -0.09 & -0.07 & -0.09 & -0.02 \\
     & model D    & -0.04 & -0.07 & -0.10 & -0.08 & -0.08 & -0.07 & -0.10 & -0.09 & -0.10 & -0.09 \\
     & $\Delta$EW & $\pm$0.12 & $\pm$0.10 & $\pm$0.15 & $\pm$0.15 & $\pm$0.15 & 
                    $\pm$0.10 & $\pm$0.08 & $\pm$0.07 & $\pm$0.08 & $\pm$0.26 \\
     & Adopted    & $^{+0.16}_{-0.13}$ & $\pm$0.12 & $^{+0.15}_{-0.18}$ & $\pm$0.17 & $\pm$0.17 &
		    $\pm$0.12 & $\pm$0.12 & $\pm$0.10 & $^{+0.10}_{-0.13}$ & $\pm$0.28 \\

4661 & model A    &  0.10 &  0.04 &  0.06 &  0.06 &  0.05 &  0.04 &  0.09 &  0.06 &  0.10 &  0.06 \\  
     & model B    &  0.12 &  0.08 &  0.08 &  0.08 &  0.10 &  0.07 &  0.09 & -0.01 &  0.09 &  0.11 \\
     & model C    & -0.03 & -0.06 & -0.06 & -0.04 & -0.02 & -0.05 & -0.11 & -0.08 & -0.09 & -0.07 \\
     & model D    & -0.05 & -0.08 & -0.08 & -0.08 & -0.07 & -0.08 & -0.11 & -0.09 & -0.10 & -0.11 \\
     & $\Delta$EW & $\pm$0.15 & $\pm$0.13 & \nodata & $\pm$0.25 & $\pm$0.24 & 
                    $\pm$0.11 & $\pm$0.08 & $\pm$0.10 & $\pm$0.08 & $\pm$0.24 \\
     & Adopted    & $^{+0.19}_{-0.16}$ & $\pm$0.15 & \nodata & $\pm$0.26 & $\pm$0.26 &
		    $\pm$0.14 & $\pm$0.13 & $\pm$0.13 & $\pm$0.13 & $\pm$0.26 \\

\enddata
\tablecomments{Uncertainties in the O\,{\sc ii} abundace determinations. 
Models A, B, C, D
refer to models calculated with the parameters from Table 3 and the following
parameter shifts in \teff, \g~ and $\xi$, respectively:
model A, (+1000 K, +0.1 dex, -2 km s$^{-1}$); 
model B, (+500 K, +0.05 dex, 0 km s$^{-1}$);
model C, (-800 K, -0.08 dex, +2 km s$^{-1}$); 
model D, (-500 K, -0.05 dex, 0 km s$^{-1}$).
$\Delta$EW refers to the uncertainty caused by the equivalent width 
error measurements.
Uncertainties have been added quadratically to give the adopted errors.
Cool and hot models of 1054 refer to the results obtained with the
Si\,{\sc iii}/Si\,{\sc iv} and Si\,{\sc ii}/Si\,{\sc iii} ionization
equilibria respectively}
\end{deluxetable}


%% file: table5.tex
\begin{deluxetable}{llccccccc}
\tablecolumns{9}
\tabletypesize{\scriptsize}
\tablecaption{Silicon abundances of the Program Stars. \label{siab}} 
\tablehead{
\multicolumn{2}{c}{M33 star} & \colhead{Si\,{\sc iii}} & \colhead{Si\,{\sc iii}}& 
\colhead{Si\,{\sc iii}} & \colhead{Si\,{\sc iv}} & \colhead{Si\,{\sc ii}} & 
\colhead{Si\,{\sc ii}} & \colhead{$12 + \log({\rm Si/H})$}\\
\colhead{} & \colhead{} & \colhead{4553} & \colhead{4568} & \colhead{4574} & 
\colhead{4116} & \colhead{4128} & \colhead{4130} & \colhead{}
} 
\startdata
110A &  0785 & 7.36$\pm$0.11 & 7.19$\pm$0.08 & 7.42$\pm$0.11 & 7.33$\pm$0.21 
              & \nodata & \nodata  & 7.32$\pm$0.15  \\
B38 & ob21-108 & 7.15$\pm$0.10 & 7.31$\pm$0.08 & 7.17$\pm$0.10 & 7.14$\pm$0.15
            	& \nodata & \nodata  & 7.24$\pm$0.12 \\
B133 & 0515   & 7.15$\pm$0.13 & 6.80$\pm$0.10 & 7.18$\pm$0.12 & 7.07$\pm$0.22    
              & \nodata & \nodata & 7.04$\pm$0.15 \\
    & 1054    & 7.58$\pm$0.15 & 7.68$\pm$0.20 & 7.61$\pm$0.15 & 7.57$\pm$0.20
              & \nodata & \nodata & 7.61$\pm$0.18 \\
    & 1054    & 7.52$\pm$0.13 & 7.61$\pm$0.18 & 7.55$\pm$0.15 & \nodata
              & 7.50$\pm$0.10 & 7.64$\pm$0.10 & 7.54$\pm$0.14 \\            
\tableline
\sidehead{Galactic  Star} 
\tableline
HD13854 &      & 7.54$\pm$0.12 & 7.57$\pm$0.11 & 7.54$\pm$0.10 & 7.68$\pm$0.11
	      & \nodata & \nodata &  7.59$\pm$0.11\\
HD14956 &      & 7.85$\pm$0.15 & 7.96$\pm$0.14 & 7.90$\pm$0.11 & 7.91$\pm$0.09 
	      & \nodata & \nodata & 7.91$\pm$0.13 \\
HD13841 &      & 7.40$\pm$0.11 & 7.48$\pm$0.11 & 7.48$\pm$0.08 & 7.53$\pm$0.18 
	      & \nodata & \nodata & 7.47$\pm$0.12 \\
HD206165&      & 7.29$\pm$0.11 & 7.31$\pm$0.10 & 7.18$\pm$0.07 & \nodata
              & 7.38$\pm$0.06 & 7.32$\pm$0.07 & 7.29$\pm$0.09 \\
HD198478&      & 7.57$\pm$0.13 & 7.50$\pm$0.12 & 7.58$\pm$0.09 & \nodata
	      & 7.53$\pm$0.05 & 7.58$\pm$0.06 & 7.55$\pm$0.09 \\
\enddata
\tablecomments{Silicon abundances are derived from
each particular line. The last column gives the stellar abundance. The first
line of M33 1054 gives the results for the \SiIII/\SiIV\, ionization
balance, and the second line gives the results for the \SiII/\SiIII\, 
ionization balance.}
\end{deluxetable}


%% file: table6.tex
\begin{deluxetable}{llccccc}
\tablecolumns{7}
\tablecaption{Oxygen Abundances of the Program Stars. \label{oab}}
\tablehead{
\multicolumn{2}{c}{M33 star} & \colhead{O\,{\sc ii}} & 
\colhead{O\,{\sc ii}}  & \colhead{O\,{\sc ii}} & \colhead{O\,{\sc ii}} & 
\colhead{$12 + \log({\rm O/H})$} \\
\colhead{} & \colhead{} & \colhead{4317} & \colhead{4319} & \colhead{4366} & 
\colhead{4661}
}
\startdata
110A &   0785 & 8.24$\pm$0.17 & 8.70$^{+0.19}_{-0.16}$ & 8.36$^{+0.16}_{-0.13}$ 
	      & 8.97$^{+0.19}_{-0.16}$ & 8.55$\pm$0.18 \\
B38 & ob21-108 & 8.47$\pm$0.12 & 8.89$\pm$0.13 & 8.46$\pm$0.12 &
               8.38$\pm$0.15 & 8.55$\pm$0.13 \\
B133 & 0515 & 7.96$\pm$0.25 & 8.34$\pm$0.17 & 8.22$\pm$0.18 & \nodata
	    & 8.22$\pm$0.20 \\
     & 1054 & 8.89$\pm$0.17 & 8.59$\pm$0.27 & 8.80$\pm$0.17 & 9.43$\pm$0.26
            & 8.90$\pm$0.22 \\
     & 1054 & 8.79$\pm$0.17 & 8.52$\pm$0.22 & 8.71$\pm$0.17 & 9.33$\pm$0.26
            & 8.79$\pm$0.20 \\
\tableline
\sidehead{Galactic  Star}
\tableline
HD13854 &    & 8.92$\pm$0.12 & 9.00$\pm$0.12 & 8.94$\pm$0.12 & 9.21$\pm$0.14	     
             & 9.00$\pm$0.13	\\
HD14956 &    & 8.96$\pm$0.13 & 9.03$\pm$0.22 & 8.65$\pm$0.12 & 8.92$\pm$0.13 
             & 8.85$\pm$0.16	\\
HD13841 &    & 8.74$^{+0.11}_{-0.14}$ & 8.81$\pm$0.11 & 8.72$\pm$0.10 	     
             & 9.05$\pm$0.13 & 8.81$\pm$0.12	\\	
HD206165&    & 8.51$\pm$0.11 & 8.46$\pm$0.17 & 8.52$^{+0.10}_{-0.13}$	     
             & 8.85$\pm$0.13 & 8.59$\pm$0.14	\\
HD198478&    & 8.63$\pm$0.32 & 8.34$\pm$0.28 & 8.25$\pm$0.28	     
             & 8.45$\pm$0.26 & 8.41$\pm$0.29\\
\enddata
\tablecomments{Oxygen abundances are derived from
each particular line. The last column gives the stellar abundance. The first
line of M33 1054 gives the results for the \SiIII/\SiIV\, ionization
balance, and the second line gives the results for the \SiII/\SiIII\, 
ionization balance.}

\end{deluxetable}


%% file: table7.tex
\begin{deluxetable}{llccccccc}
\tablecolumns{9}
\tabletypesize{\scriptsize}
\tablecaption{Silicon Differential Abundances of the Program Stars. 
\label{difsiab}}
\tablehead{
\multicolumn{2}{c}{M33 star} & \colhead{Galactic} & \colhead{Si\,{\sc iii}} & 
\colhead{Si\,{\sc iii}}  & \colhead{Si\,{\sc iii}}& \colhead{Si\,{\sc iv}} & 
\colhead{Si\,{\sc ii}} & \colhead{Si\,{\sc ii}} \\
\colhead{} & \colhead{} & \colhead{Standard} & \colhead{4553} &
\colhead{4568} & \colhead{4574} & \colhead{4116} & \colhead{4128} & 
\colhead{4130}
} 
\startdata
110A & 0785 & HD13854  & $-0.18\pm$0.16 & $-0.38\pm$0.14 
                       & $-0.12\pm$0.15 & $-0.35\pm$0.24 & \nodata & \nodata\\  
     &      & HD14956  & $-0.49\pm$0.18 & $-0.77\pm$0.16 
                       & $-0.48\pm$0.15 & $-0.58\pm$0.23 & \nodata & \nodata\\
B38 & ob21-108& HD13854& $-0.39\pm$0.16 & $-0.26\pm$0.16 
                       & $-0.37\pm$0.15 & $-0.54\pm$0.17 & \nodata & \nodata\\
    &       & HD14956  & $-0.70\pm$0.18 & $-0.65\pm$0.16 
		       & $-0.73\pm$0.15 & $-0.77\pm$0.17 & \nodata & \nodata\\
B133 & 0515 & HD13854  & $-0.39\pm$0.18 & $-0.77\pm$0.15 
		       & $-0.37\pm$0.15 & $-0.61\pm$0.25 & \nodata & \nodata\\
     &      & HD13841  & $-0.25\pm$0.17 & $-0.68\pm$0.15 
		       & $-0.30\pm$0.14 & $-0.46\pm$0.28 & \nodata & \nodata\\
    & 1054  & HD13841  & +0.18$\pm$0.18 & +0.20$\pm$0.23 
		       & +0.13$\pm$0.17 & 0.04$\pm$0.25  & \nodata & \nodata\\
    &       & HD206165 & +0.23$\pm$0.17 & +0.30$\pm$0.20 
		  & +0.37$\pm$0.16 & \nodata & +0.12$\pm$0.11 & +0.32$\pm$0.12\\
\enddata
\tablecomments{Silicon differential abundances are
derived from each particular line. The first
line of M33 1054 gives the results for the \SiIII/\SiIV\, ionization
balance, and the second line gives the results for the \SiII/\SiIII\, 
ionization balance.}
\end{deluxetable}


%% file: table8.tex
\begin{deluxetable}{llccccc}
\tablecolumns{7}
\tablecaption{Oxygen Differential Abundances of the Program Stars. 
\label{difoab}}
\tablehead{
\multicolumn{2}{c}{~M33 star} & \colhead{Galactic} & 
\colhead{O\,{\sc ii}} & \colhead{O\,{\sc ii}}  & \colhead{O\,{\sc ii}} & 
\colhead{O\,{\sc ii}} \\
\colhead{} & \colhead{} & \colhead{Standard} & \colhead{4317} &
\colhead{4319} & \colhead{4366} & \colhead{4661} \\
}
\startdata
110A & 0785     & HD13854 & $-0.68\pm$0.21 & $-0.30^{+0.22}_{-0.20}$ & 
			    $-0.58^{+0.20}_{-0.18}$ & $-0.22^{+0.24}_{-0.21}$ \\
     &          & HD14956 & $-0.72\pm$0.21 & $-0.33^{+0.29}_{-0.27}$ & 
			    $-0.29^{+0.20}_{-0.18}$ & $+0.05^{+0.23}_{-0.21}$ \\
                     
B38 & ob21-108  & HD13854 & $-0.45\pm$0.17 & $-0.11\pm$0.18 & $-0.48\pm$0.17 
                          & $-0.83\pm$0.21 \\
    &           & HD14956 & $-0.49\pm$0.18 & $-0.14\pm$0.26 & $-0.19\pm$0.17 
                          & $-0.54\pm$0.21 \\
                           
B133 & 0515     & HD13854 & $-0.96\pm$0.28 & $-0.66\pm$0.21 & $-0.72\pm$0.22
                          & \nodata \\
     &          & HD13841 & $-0.78^{+0.27}_{-0.29}$ & $-0.47\pm$0.20 & $-0.50\pm$0.21
                          & \nodata \\

    & 1054      & HD13841 & $+0.15^{+0.20}_{-0.22}$ & $-0.22\pm$0.29 & +0.08$\pm$0.20
                          & $+0.38\pm$0.29 \\
                           
    &           & HD206165& +0.28$\pm$0.20 & +0.06$\pm$0.28 & +0.19$\pm$0.21
			  & $+0.48\pm$0.29 \\                          
\enddata
\tablecomments{Oxygen differential abundances are
derived from each particular line. The first
line of M33 1054 gives the results for the \SiIII/\SiIV\, ionization
balance, and the second line gives the results for the \SiII/\SiIII\, 
ionization balance}
\end{deluxetable}


%% file: table9.tex
\begin{deluxetable}{lcccc}
\tablecolumns{5}
\tablecaption{Si and O Non-LTE Differential Abundances. \label{grads}} 
\tablehead{
\colhead{M33 star} &  \colhead{$\rho$}  &  \colhead{Standard}  &  
\colhead{$\log({\rm Si/H})_{\rm M33}-\log({\rm Si/H})_{\rm Gal}$} &  
\colhead{$\log({\rm O/H})_{\rm M33}-\log({\rm O/H})_{\rm Gal}$}
}
\startdata
1054     &  1$'$.0  & HD206165   & +0.25$\pm$0.16   & +0.25$\pm$0.25  \\
	 &          & HD13841    & +0.14$\pm$0.21   & +0.18$\pm$0.25 \\
	 &          & Average    & +0.20$\pm$0.19   & +0.22$\pm$0.25 \\
B133     &  9$'$.4  & HD13854    & $-0.53\pm$0.19   & $-0.75\pm$0.24   \\
	 &          & HD13841    & $-0.42\pm$0.19   & $-0.55\pm$0.24  \\
	 &          & Average    & $-0.48\pm$0.19   & $-0.65\pm$0.24  \\
ob21-108 &  11$'$.7 & HD14956    & $-0.71\pm$0.17   & $-0.35\pm$0.21 \\
         &          & HD13854    & $-0.39\pm$0.16   & $-0.44\pm$0.18 \\
         &          & Average    & $-0.54\pm$0.17   & $-0.40\pm$0.20 \\
110A     &  17$'$.1 & HD14956    & $-0.58\pm$0.18   & $-0.33\pm$0.23   \\
         &          & HD13854    & $-0.25\pm$0.18   & $-0.47\pm$0.21   \\
         &          & Average    & $-0.41\pm$0.18   & $-0.41\pm$0.22 \\
\enddata
\tablecomments{Silicon and oxygen non-LTE differential abundances for 
the M33  B supergiants compared to Galactic standards. $\rho$ is the 
galactocentric distance in arcminutes.} 
\end{deluxetable}


%% file: figus.tex
\begin{figure}
\label{spec}
\plotone{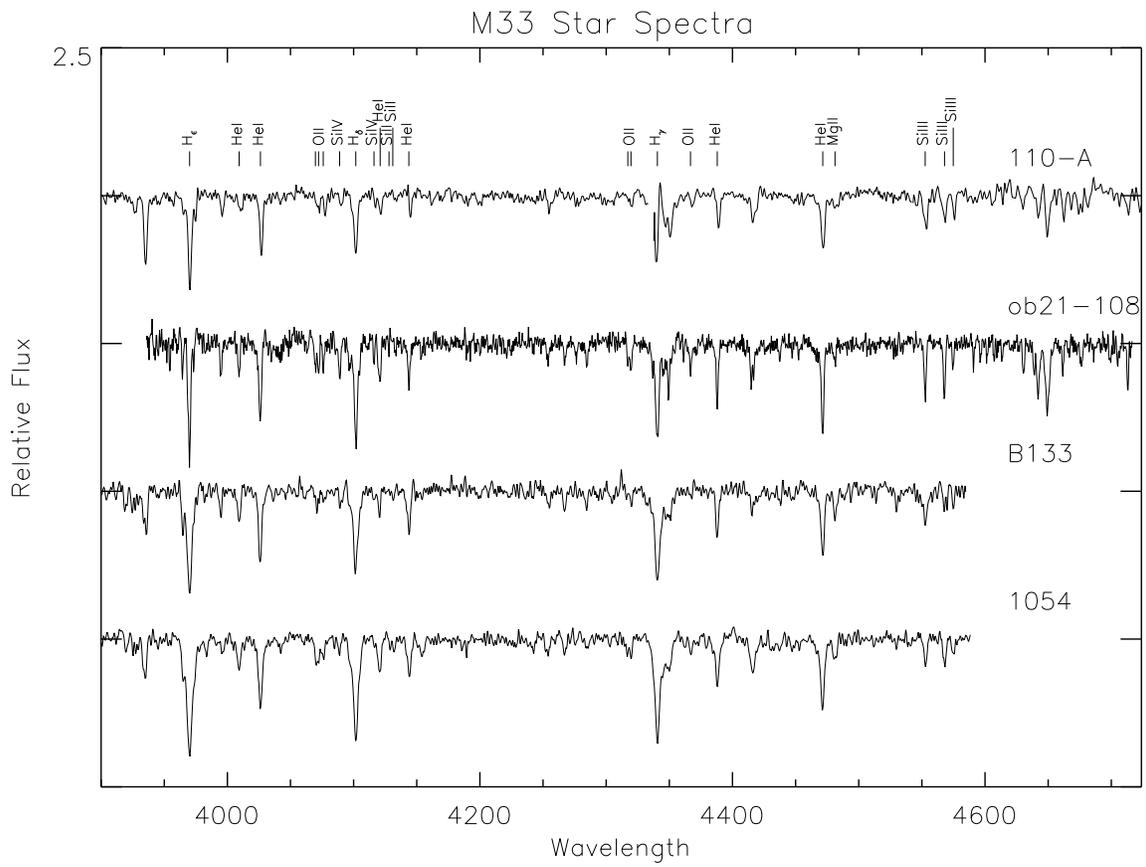}
\caption{Blue spectra of the M33 stars analyzed here, 
with the most
important lines for this paper marked. The spectrum for 110-A is a
merging of two different observing campaigns, described in Herrero et al.
(1994). The gap in the blue wing of \Hg\, in this star 
is due to a cosmic ray.}
\end{figure}

\clearpage

\begin{figure}
\label{haspec}
\plotone{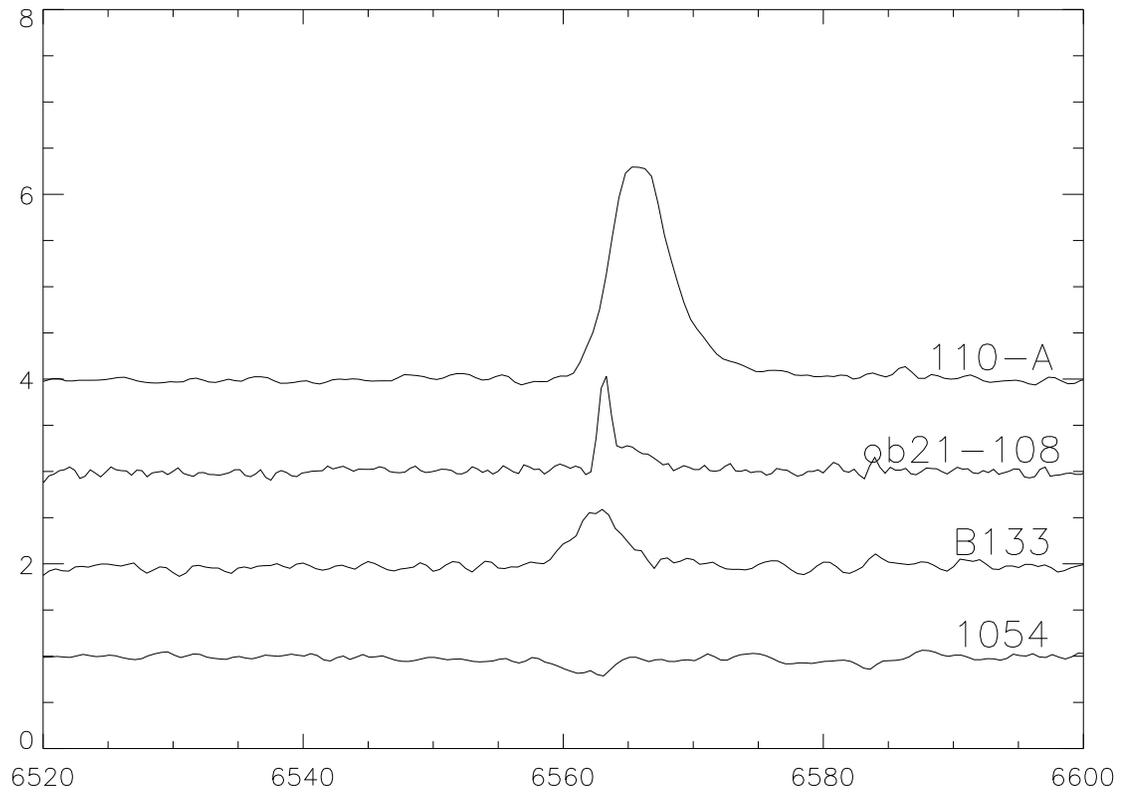}
\caption{\Ha\, spectra of the M33 stars analyzed here.}
\end{figure}

\clearpage

\begin{figure}
\label{noly}
\plotone{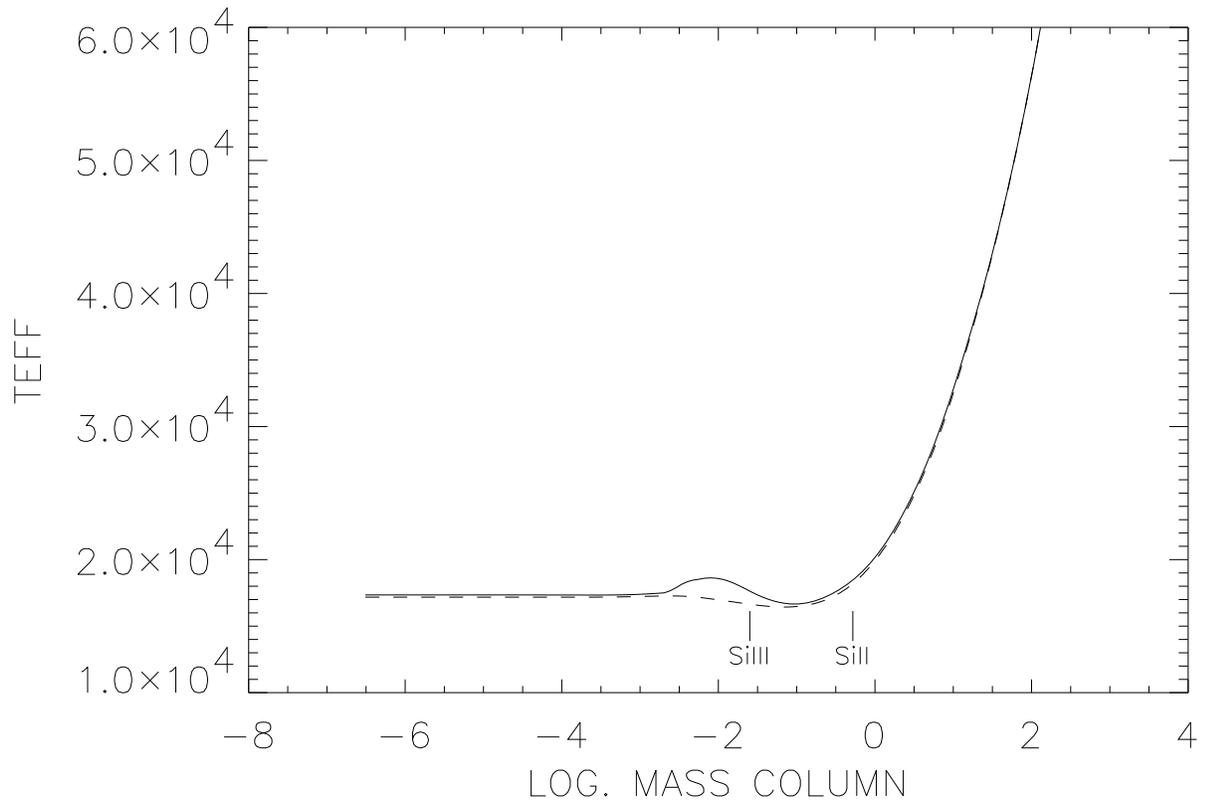}
\caption{Comparison of the temperature runs of two models at 
\teff\ = 22\,500K, \g\ = 2.50, \eps\ = 0.09, with the Lyman lines in detailed
balance (dashed line) and not in detailed balance (solid line)
during the model atmosphere calculations.}
\end{figure}

\clearpage

\begin{figure}
\label{noly2}
\plotone{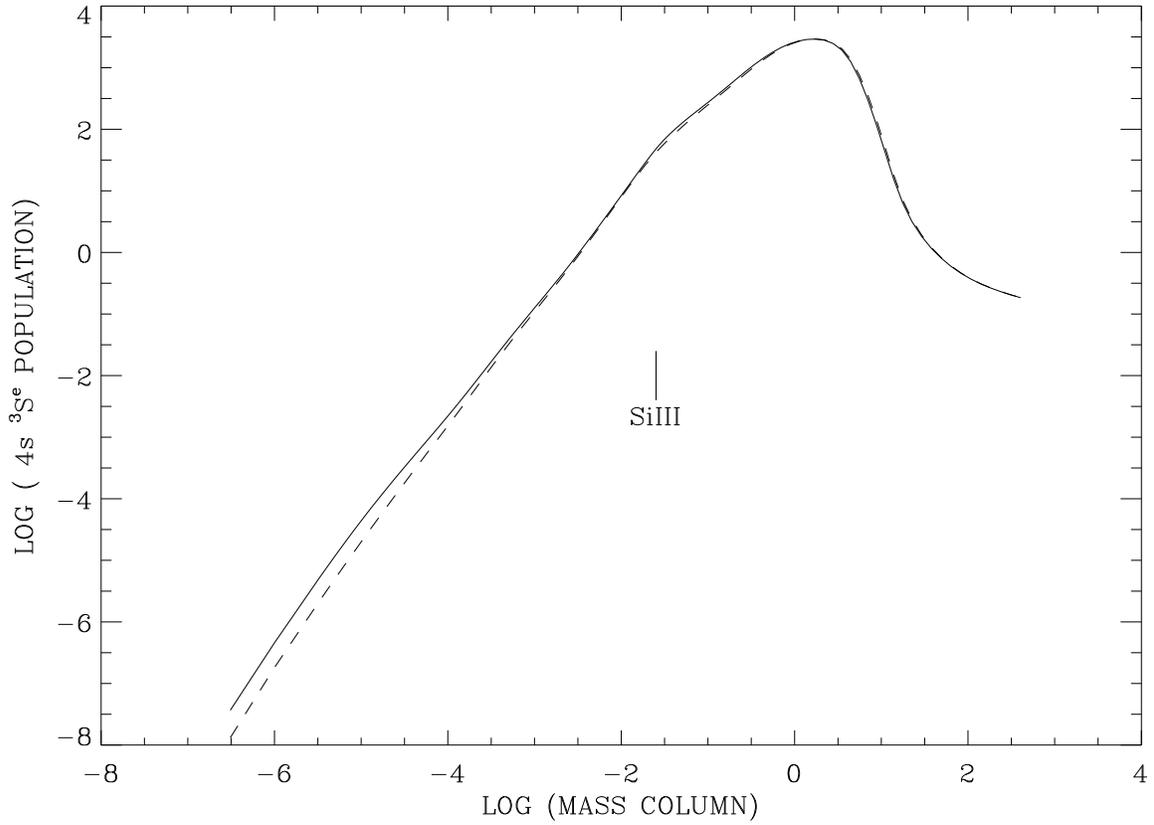}
\caption{Comparison of the populations of 4s $^3$S$^e$, 
the lower level
of \SiIII~$\lambda$4553, in two models at 
\teff\ = 22\,500K, \g\ = 2.50, \eps\ = 0.09, with the Lyman lines in detailed
balance (dashed line) and not in detailed balance (solid line)
during the model atmosphere calculations. The formation depth of the
line core is also marked. Mass column is given in g cm$^{-2}$ and
populations in atoms cm$^{-3}$.}
\end{figure}

\clearpage

\begin{figure}
\label{nores1}
\plotone{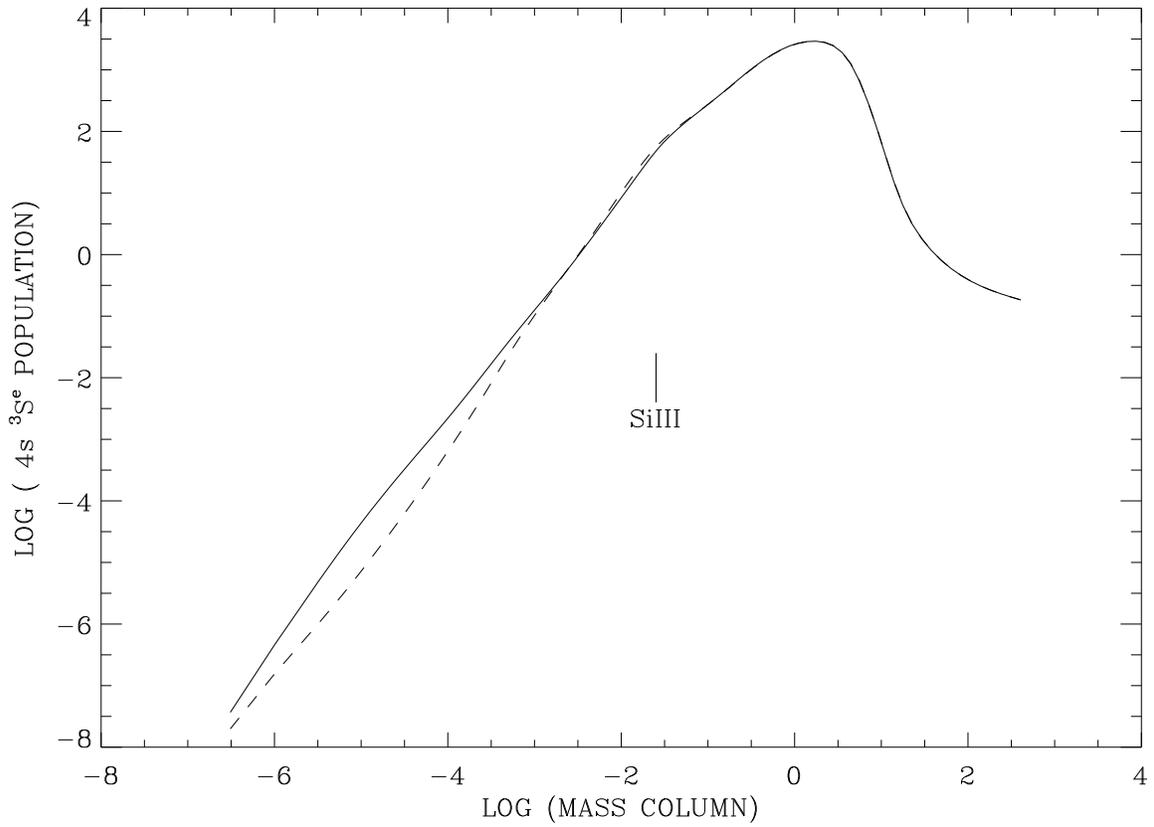}
\figcaption{Comparison of the populations of 4s $^3$S$^e$, 
the lower level
of \SiIII~$\lambda$4553, in two models at 
\teff\ = 22\,500K, \g\ = 2.50, \eps\ = 0.09, with the \SiIII~ resonance
lines in detailed balance (dashed line) and not in detailed balance 
(solid line) during the line formation calculations, with
no microturbulent velocity. The formation depth of the
line core is also marked. Mass column is given in g cm$^{-2}$ and
populations in atoms cm$^{-3}$.}
\end{figure}

\clearpage

\begin{figure}
\label{nores2}
\plotone{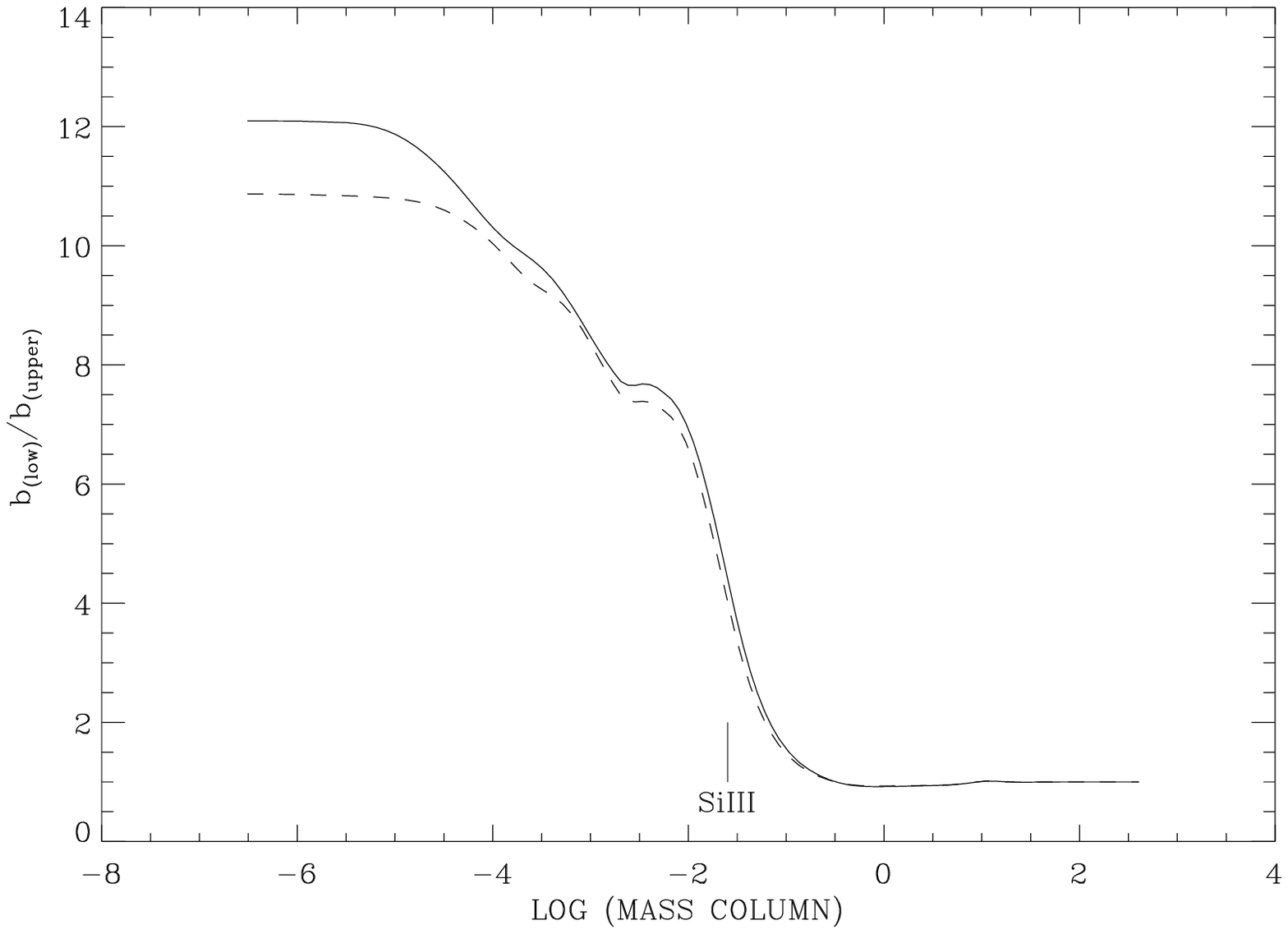}
\caption{Comparison of the ratio of the departure 
coefficients of
4s $^3$S$^e$ and 4p $^3$P$^o$, the lower and upper levels 
of \SiIII~$\lambda\lambda$4553, 4568, 4574 in two models at 
\teff\ = 22\,500K, \g\ = 2.50, \eps\ = 0.09, with the \SiIII~ resonance
lines in detailed balance (dashed line) and not in detailed balance 
(solid line) during the line formation calculations. 
The formation depth of the
line cores is also marked. Mass column is given in g cm$^{-2}$. Note
that this time the vertical axis is not logarithmic as in the previous
figures.}
\end{figure}

\clearpage 

\begin{figure}
\label{nores_15}
\plotone{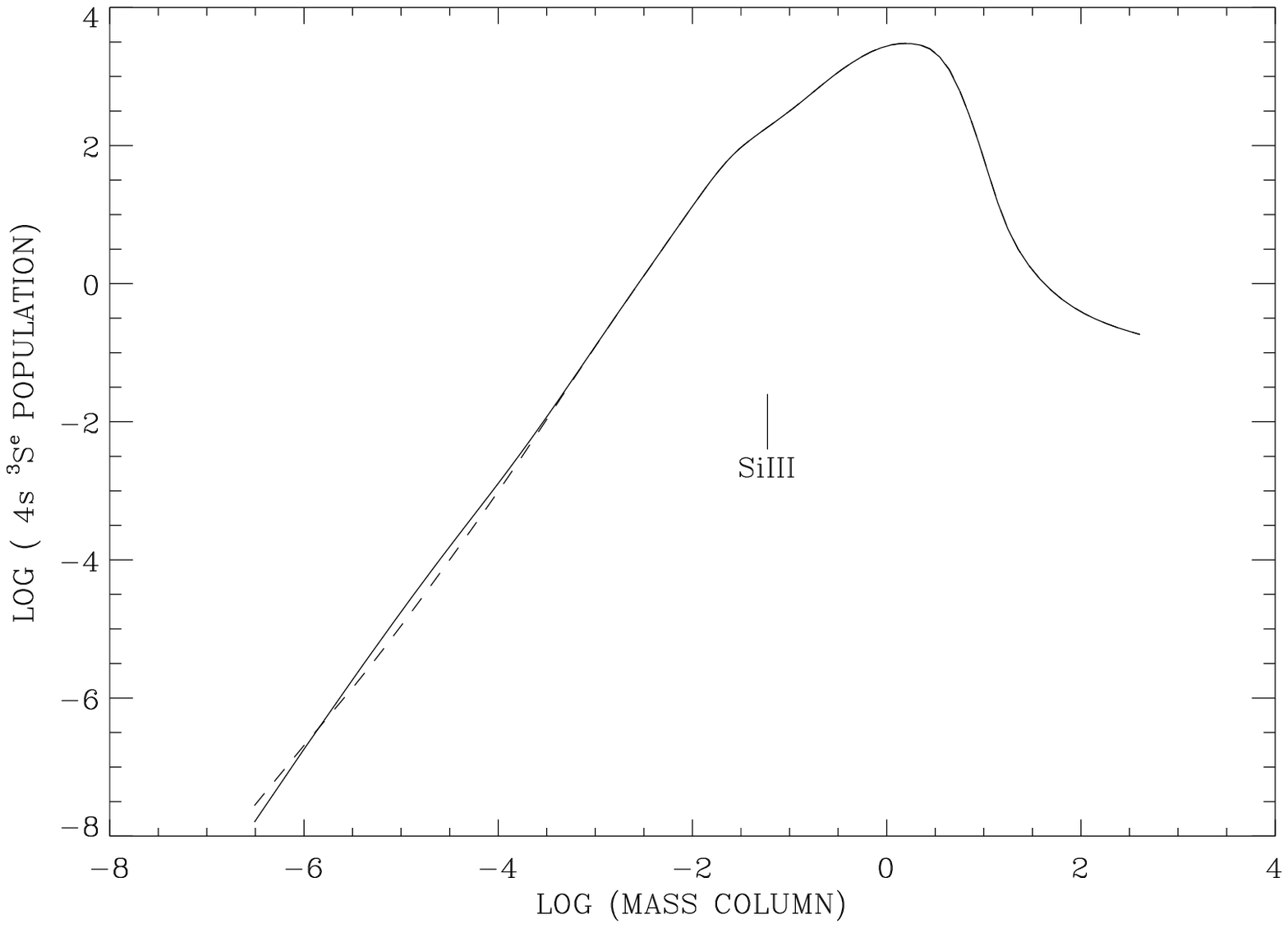}
\caption{Comparison of the populations of 4s $^3$S$^e$, 
the lower level
of \SiIII~$\lambda$4553, in two models at 
\teff\ = 22\,500K, \g\ = 2.50, \eps\ = 0.09, with the \SiIII~ resonance
lines in detailed balance (dashed line) and not in detailed balance 
(solid line) during the line formation calculations, with a microturbulent
velocity of 15 km s$^{-1}$. Compare with Figure 5 where
no microturbulence was included. The formation depth of the
line core is also marked. Mass column is given in g cm$^{-2}$ and
populations in atoms cm$^{-3}$.}
\end{figure}

\clearpage

\begin{figure}
\label{fproc1}
\plotone{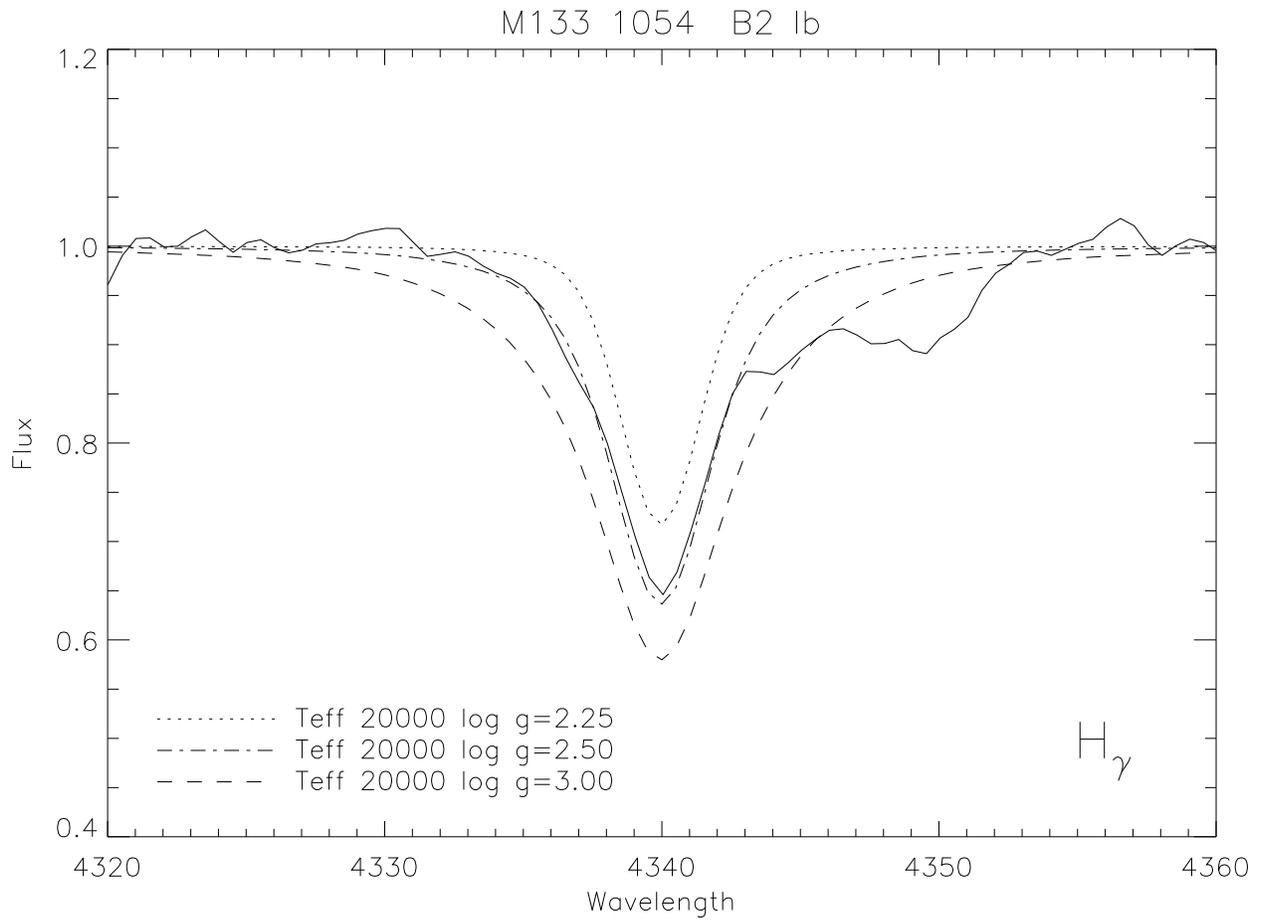}
\caption{The determination of gravity at 
\teff\ = 20\,000K in M33 1054 
using the wings of \Hg. The same procedure is used to determine
it at other temperatures, giving possible pairs of (\teff , \g).}
\end{figure}

\clearpage

\begin{figure}
\label{fproc2}
\plotone{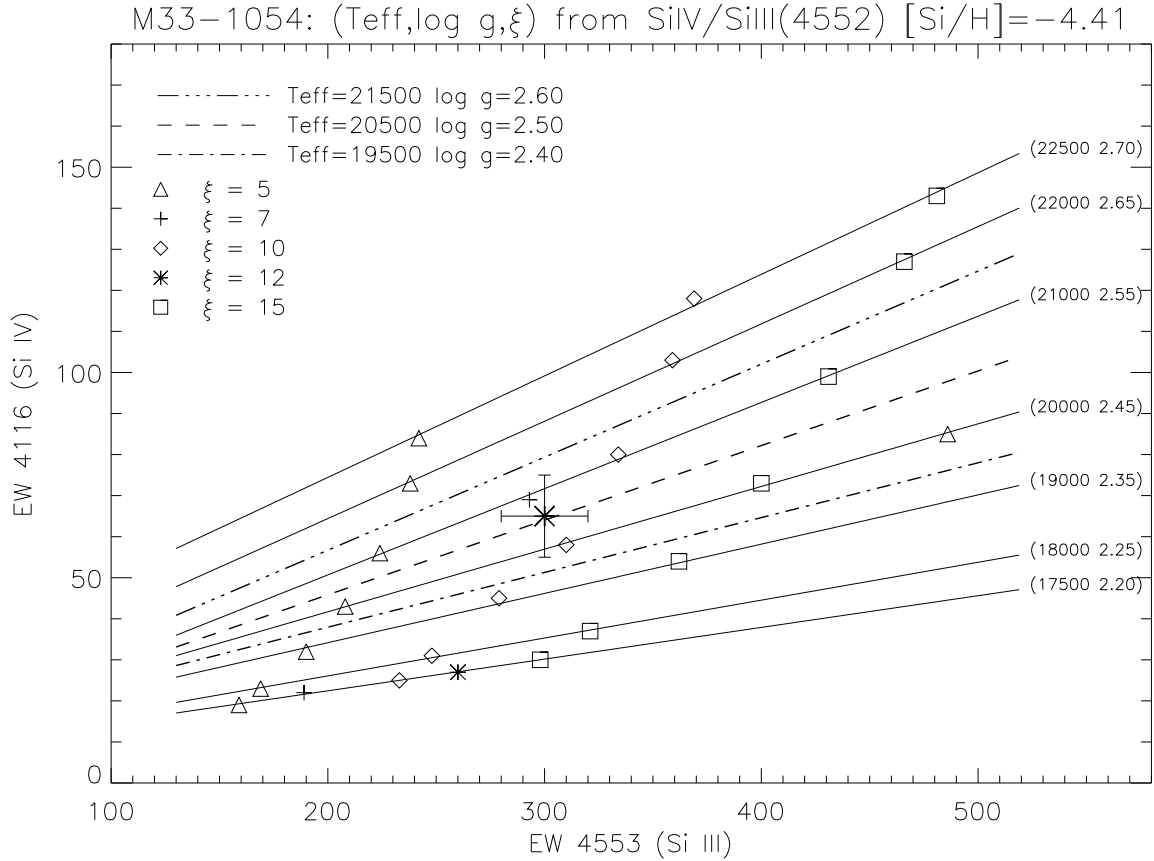}
\caption{Joint determination of microturbulence, 
\teff~ and \g~ 
at solar Si abundance for \SiIII\,$\lambda$4553 and \SiIV\,$\lambda$4116.
The possible pairs of \teff~ and \g~ were
fixed by the Balmer lines wings. Solid lines indicate models
that were explicitly calculated. Different symbols are used for
different microturbulent velocities, as indicated in the figure
(values are given in km s$^{-1}$).
The cross marks the observed equivalent width. The dashed,
dash-dotted and dash-double dotted lines have been obtained 
by interpolation and give, respectively, the model parameters
that fit the observed equivalent width and their 
adopted upper and lower limits. Similar diagrams are obtained
for other pairs of Si lines, to finally obtain a value
for \teff, \g, and $\xi$ at the assumed Si abundance.}
\end{figure}

\clearpage

\begin{figure}
\label{scale}
\plotone{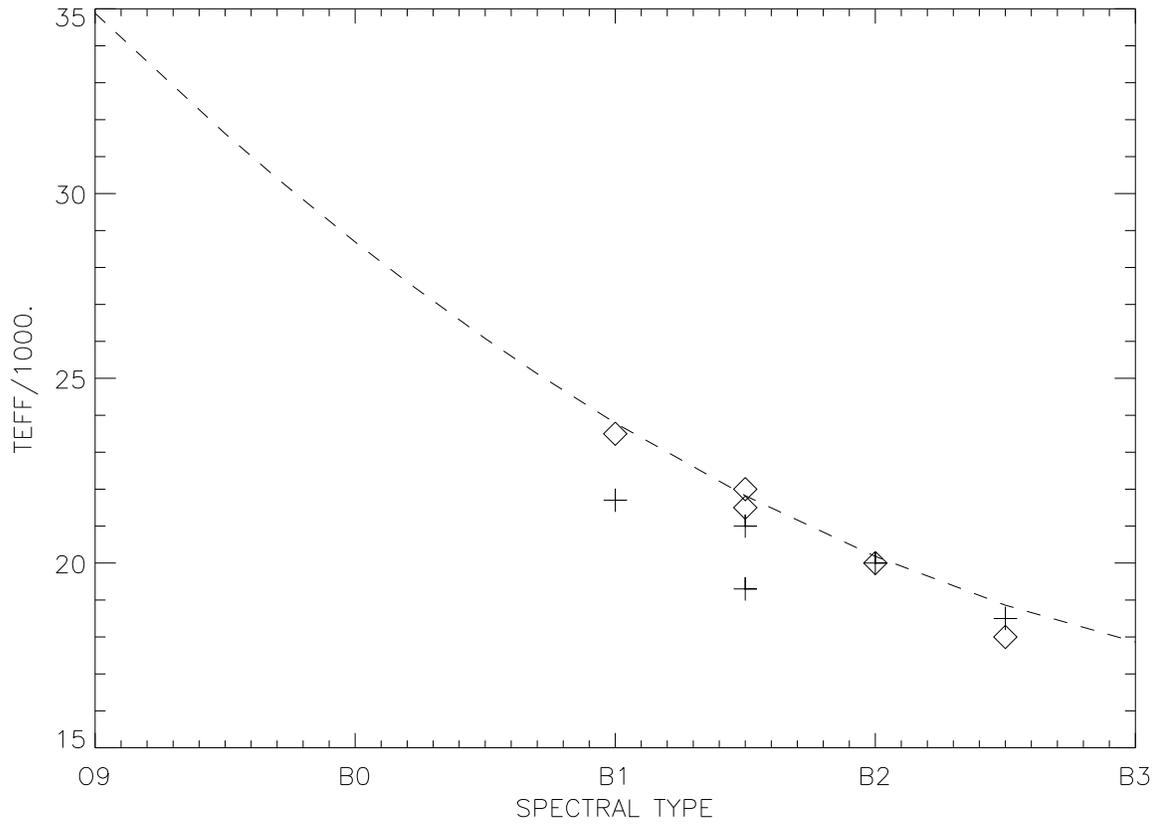}
\caption{The temperature scale for B supergiants from the 
fit to the
data by McErlean et al. (1998a). Diamonds represent the individual
values obtained by McErlean et al. (1988a) for the Galactic
stars analyzed here, and crosses represent the values obtained in the present
work for the same objects.}
\end{figure}

\clearpage

\begin{figure}
\label{bin_blue}
\plotone{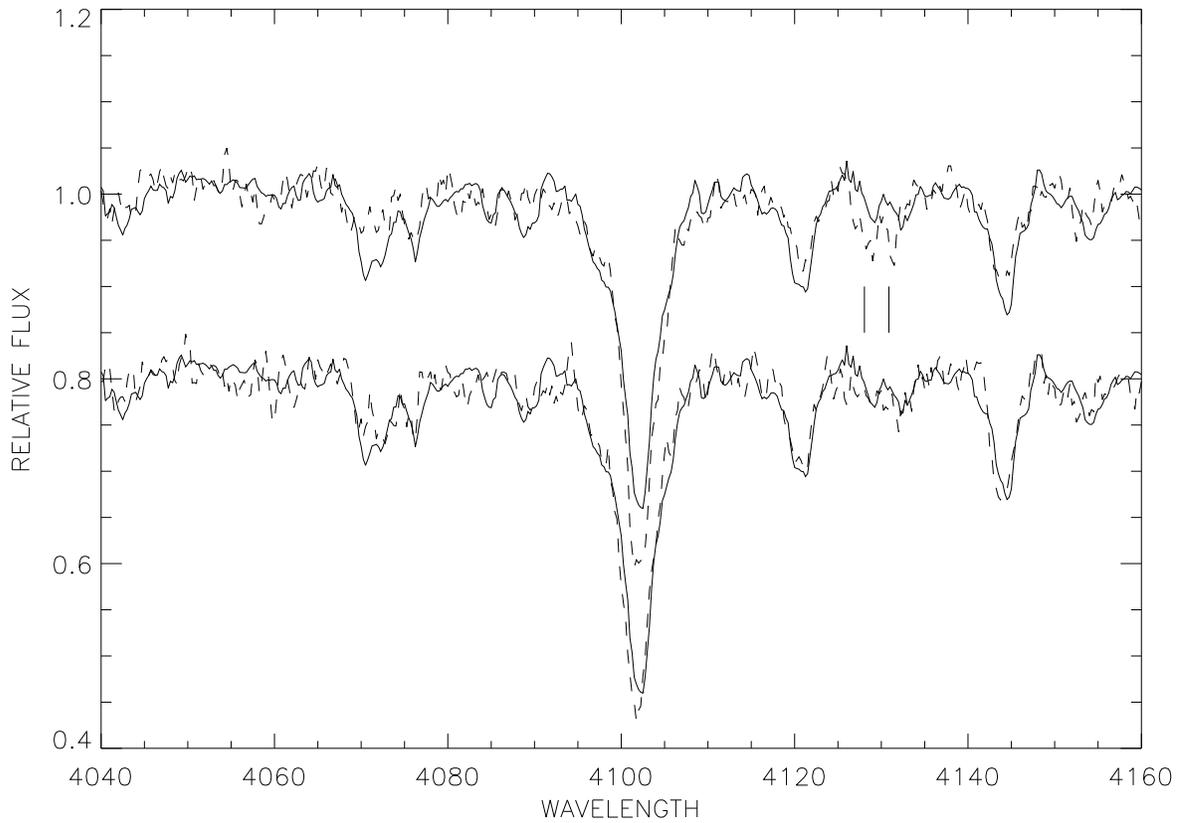}
\caption{Comparison of a portion of the blue 
spectrum of M33 1054
(solid line) with the simulations of a (B2Ib+A0Ib) binary (dashed line,
above) and a single B2Ib star (dashed line, displaced 0.2 to
the bottom for clarity). The vertical lines between the spectra
mark the laboratory positions of \SiII\,$\lambda\lambda$4128, 4130.
We see that the simulated binary has 
more conspicuous \SiII\, lines, and that the single B2Ib star
can simulate the apparent displacement of these lines because of
noise and low S/N.}
\end{figure}

\clearpage

\begin{figure}
\label{bin_red}
\plotone{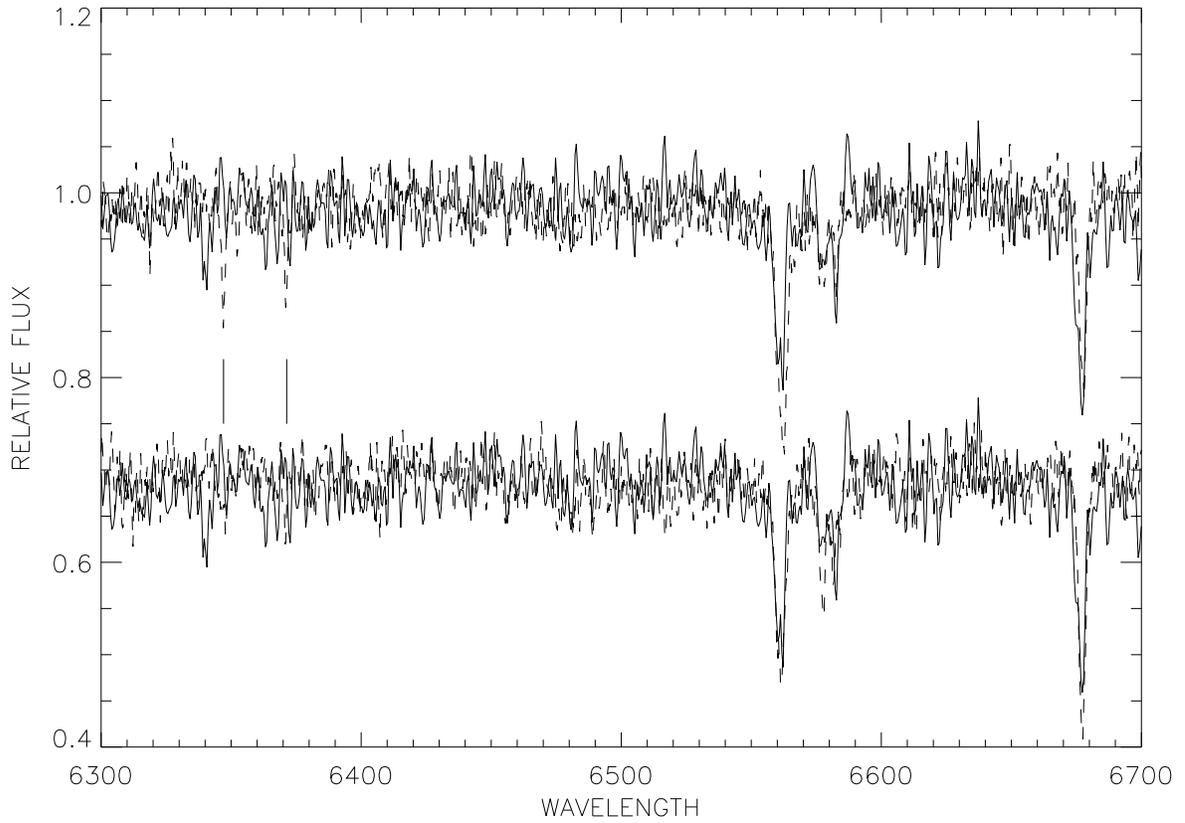}
\figcaption{Comparison of a portion of the red 
spectrum of M33 1054
(solid line) with the simulations of a (B2Ib+A0Ib) binary (dashed line,
above) and a single B2Ib star (dashed line, displaced 0.3 to
the bottom for clarity). The vertical lines between the spectra
mark the laboratory positions of \SiII\,$\lambda\lambda$6347, 6371.
We would expect to see the strong
\SiII\ lines if M33 1054 were actually a binary of the
assumed characteristics.}
\end{figure}

\clearpage

\begin{figure}
\label{1054fit}
\plotone{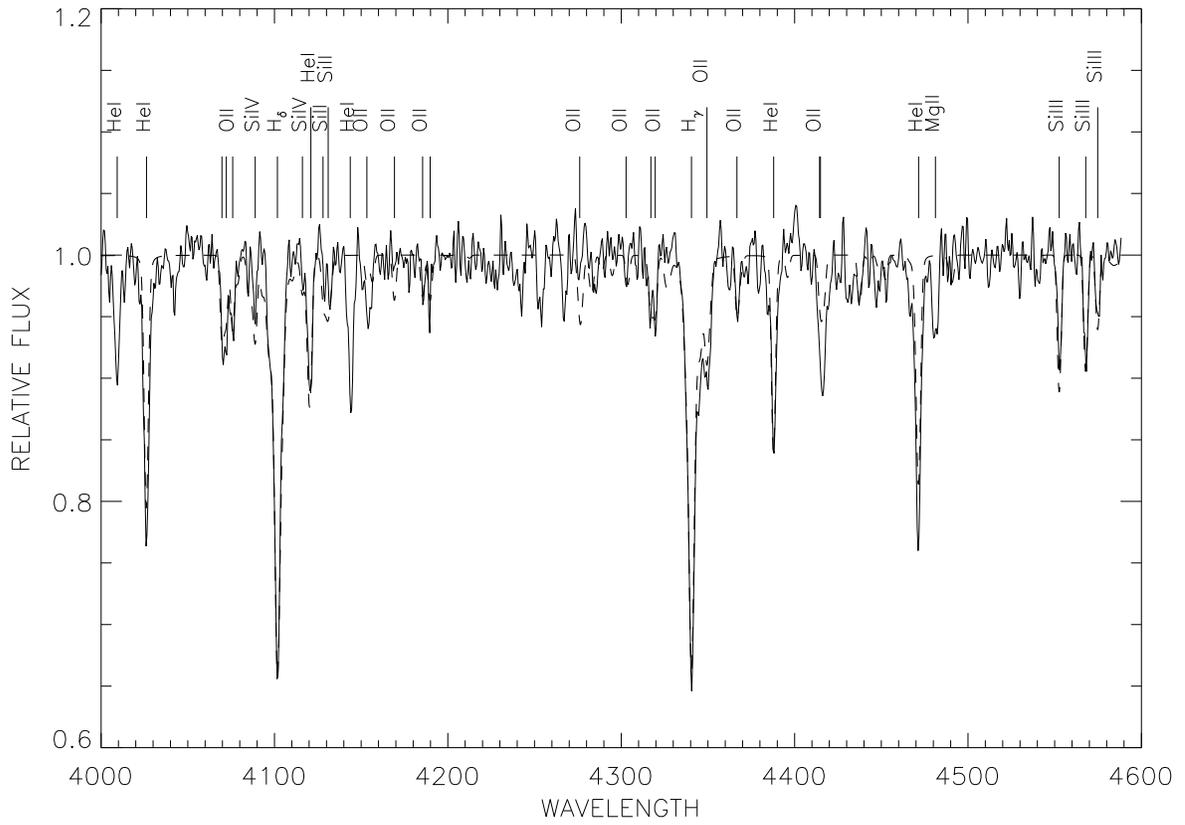}
\figcaption{The fit to the observed spectrum of M33 1054. 
We have marked the same lines as in Figure 1 and also those calculated
with SURFACE.}
\end{figure}

\clearpage

\begin{figure}
\label{ograd}
\plotone{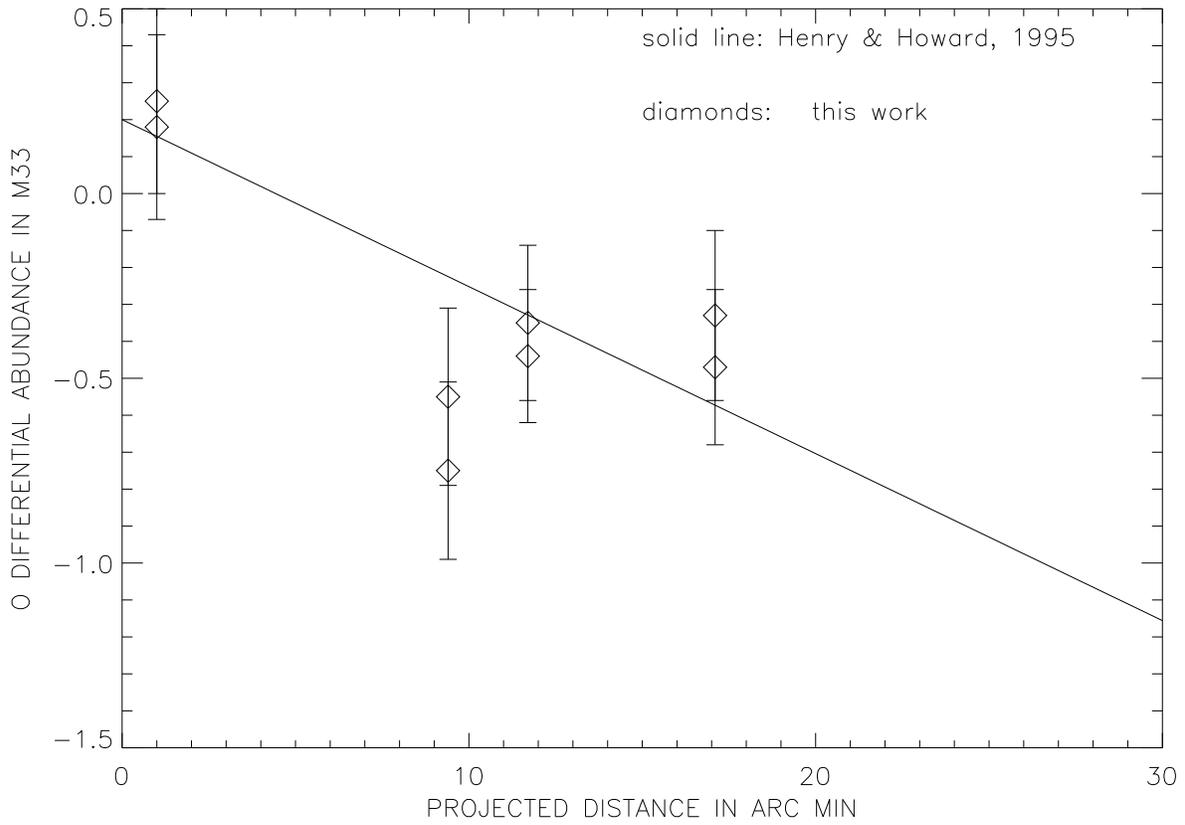}
\figcaption{The O abundance gradient in M33 from stellar data. 
Symbols with the same abscissa
values refer to the same object. The line represents the linear gradient
derived by Henry \& Howard (1995) from \HII\, region data taken
from different authors.}
\end{figure}

\clearpage

\begin{figure}
\label{ohflat}
\plotone{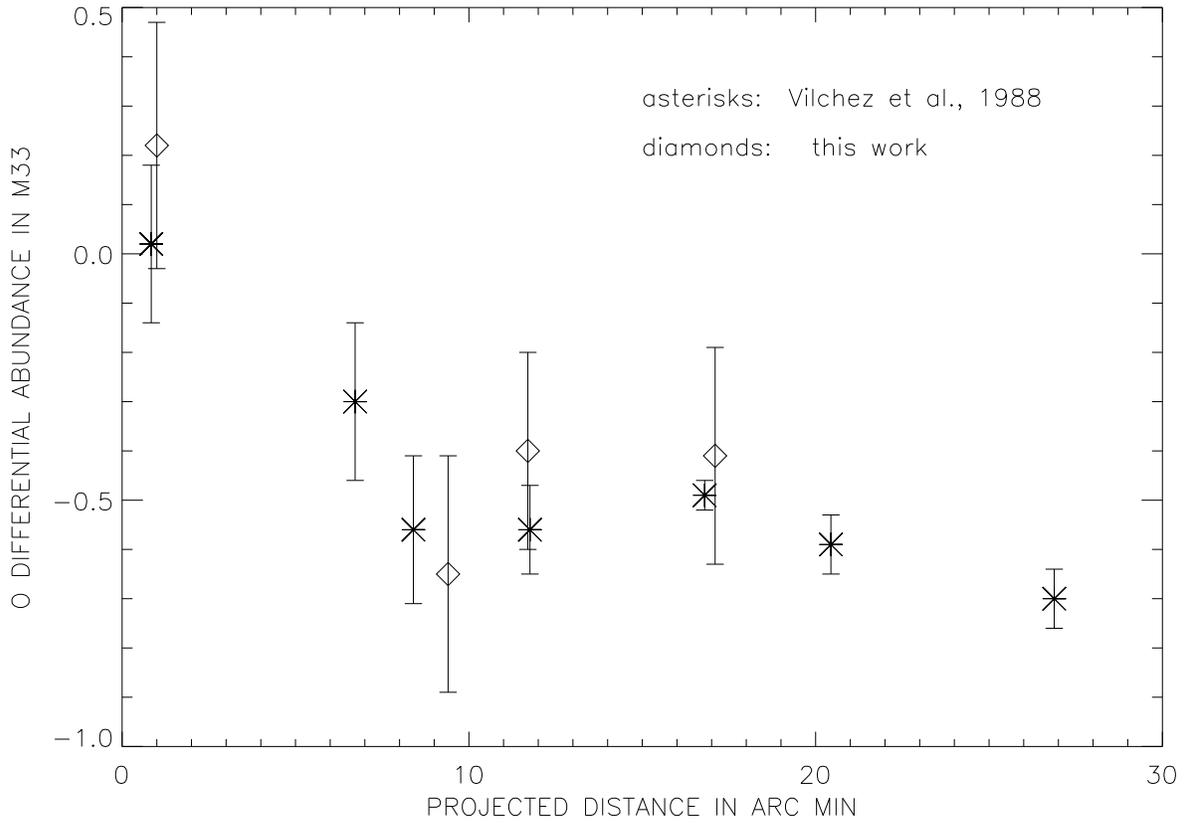}
\figcaption{The differential abundance gradient in M33 
for stars in this
work (diamonds) and from \HII\, regions (Vilchez et al. (1988))
We have used the average values given in Table 9 for clarity.}
\end{figure}

\clearpage

\begin{figure}
\label{sigrad}
\plotone{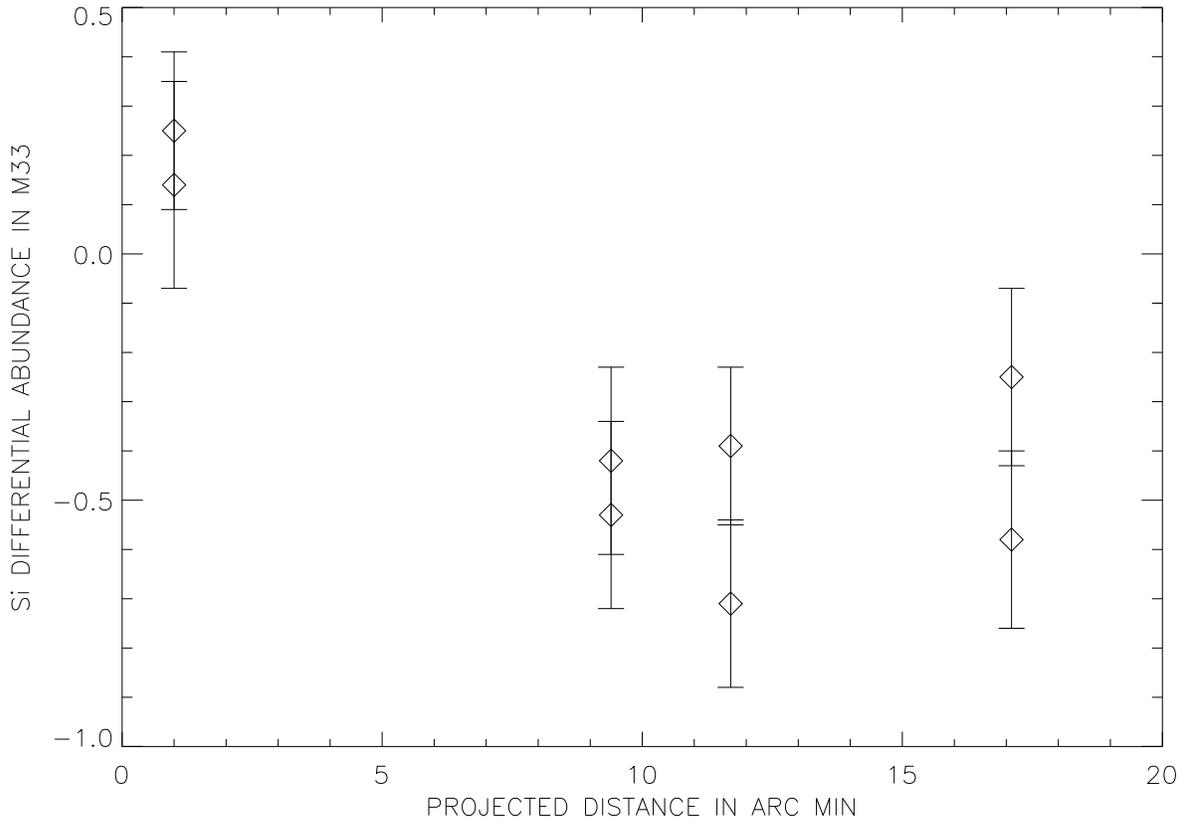}
\figcaption{The Si abundance gradient in M33 from stellar data. 
Symbols with the same abscissa values refer to the same object.}
\end{figure}

\clearpage

\begin{figure}
\label{siocorr}
\plotone{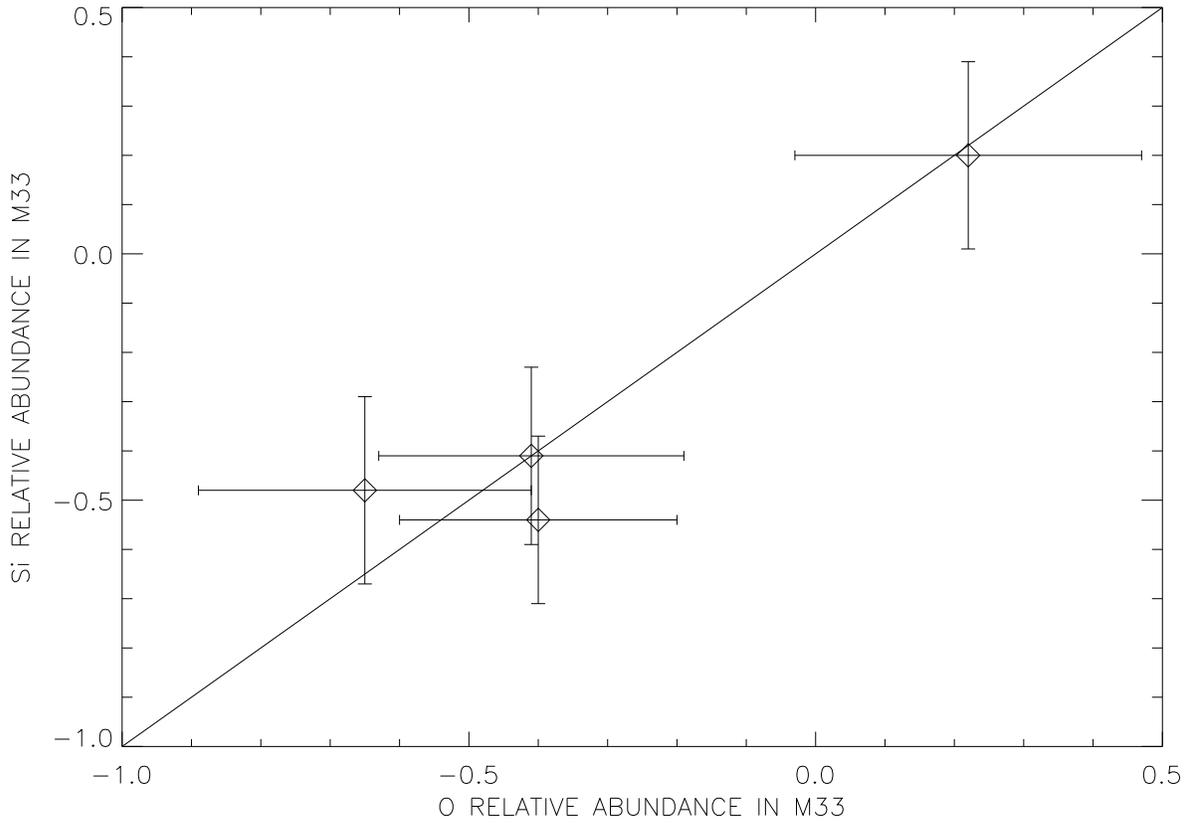}
\caption{The correlation of the Si and O relative abundances in
stars of M33 with respect to Galactic stars. The average values of 
for each individual object have been used for clarity.}
\end{figure}

\clearpage

\begin{figure}
\label{orelat}
\plotone{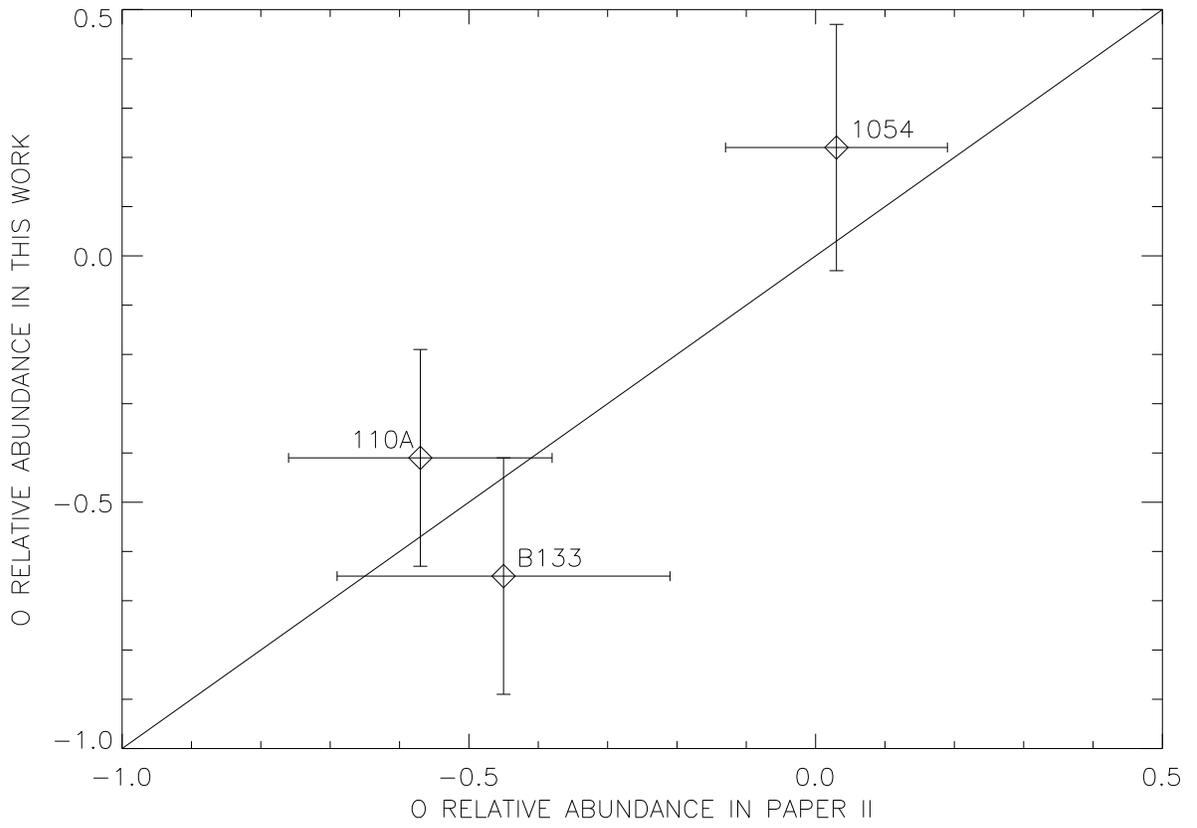}
\figcaption{The relative O abundances obtained in Paper II
and in this paper, forthe three stars in common.}
\end{figure}